\documentclass[iop]{emulateapj}
\usepackage{rotating}
\usepackage{color} 
\usepackage{amsmath}
\newcommand{\kms}       {\mbox{km s$^{-1}$}}%
\newcommand{\msun}      {\mbox{$M_\odot$}}%
\newcommand{\cur}      {\mbox{$u-r$}}%
\newcommand{\cuj}      {\mbox{$u-J$}}%
\newcommand{\cuk}      {\mbox{$u-K$}}%
\newcommand{\cgr}      {\mbox{$g-r$}}%
\newcommand{\cnuvr}      {\mbox{NUV$-r$}}%
\newcommand{\curm}      {\mbox{$(u-r)^m$}}%
\newcommand{\cujm}      {\mbox{$(u-J)^m$}}%

\shortauthors{Eckert et al.}

\shorttitle{RESOLVE Photometry}

\begin{document}
 \title{RESOLVE Survey Photometry and Volume-limited Calibration of the Photometric Gas Fractions Technique}

\author{Kathleen D.\ Eckert\altaffilmark{1},
        Sheila J.\ Kannappan\altaffilmark{1},
        David V.\ Stark\altaffilmark{1},
        Amanda J.\ Moffett\altaffilmark{1}\altaffilmark{2},
        Mark A.\ Norris\altaffilmark{1}\altaffilmark{3},
        Elaine M.\ Snyder\altaffilmark{1}, and
        Erik A.\ Hoversten\altaffilmark{1}}

\altaffiltext{1}{Department of Physics and Astronomy, University of
  North Carolina, 141 Chapman Hall CB 3255, Chapel Hill, NC 27599,
  USA; keckert@physics.unc.edu} 

\altaffiltext{2}{International Centre
  for Radio Astronomy Research (ICRAR), The University of Western
  Australia, 35 Stirling Highway, Crawley, WA 6009, Australia}

\altaffiltext{3}{Max-Plank-Institut F\"ur Astronomie (MPIA),
  K\"onigstuhl 17, 69117 Heidelberg, Deutschland}

\begin{abstract}

We present custom-processed ultraviolet, optical, and near-infrared
photometry for the RESOLVE (REsolved Spectroscopy of a Local VolumE)
survey, a volume-limited census of stellar, gas, and dynamical mass
within two subvolumes of the nearby universe (RESOLVE-A and
RESOLVE-B). RESOLVE is complete down to baryonic mass
$\sim10^{9.1-9.3}$ \msun, probing the upper end of the dwarf galaxy
regime. In contrast to standard pipeline photometry (e.g., SDSS), our
photometry uses optimal background subtraction, avoids suppressing
color gradients, and employs multiple flux extrapolation routines to
estimate systematic errors. With these improvements, we measure
brighter magnitudes, larger radii, bluer colors, and a real increase
in scatter around the red sequence.  Combining stellar mass estimates
based on our optimized photometry with the nearly complete HI mass
census for RESOLVE-A, we create new z=0 volume-limited calibrations of
the photometric gas fractions (PGF) technique, which predicts
gas-to-stellar mass ratios (G/S) from galaxy colors and optional
additional parameters.  We analyze G/S-color residuals vs.\ potential
third parameters, finding that axial ratio is the best independent and
physically meaningful third parameter. We define a ``modified color''
from planar fits to G/S as a function of both color and axial
ratio. In the complete galaxy population, upper limits on G/S bias
linear and planar fits. We therefore model the entire PGF probability
density field, enabling iterative statistical modeling of upper limits
and prediction of full G/S probability distributions for individual
galaxies. These distributions have two-component structure in the red
color regime. Finally, we use the RESOLVE-B 21cm census to test
several PGF calibrations, finding that most systematically under- or
overestimate gas masses, but the full probability density method
performs well.
\end{abstract}


%%%%%%%%%%%%%%%%%%%%INTRODUCTION%%%%%%%%%%%%%%%%%%%%%%%%%%%

\section{Introduction}
\label{sec:intro}

%\subsection{Importance of using photometric estimators for gas}

%    A. HI measurements can be time consuming - especially when you want to know gas masses for a survey

%    B. As we move into realm of large surveys - want to be able to predict gas

%    C. Important to know the Baryonic mass of galaxies especially probing down into dwarf regime where mstars may not dominate.

As imaging surveys provide ever more sky-coverage and greater depth,
we are producing larger galaxy data sets probing to lower
masses. Photometry from these imaging surveys allows estimation of
stellar masses for galaxies, which only provides a partial view of
galaxy mass without any cold gas data.  The cold neutral gas mass
probed by 21cm atomic hydrogen (HI) observations is generally the most
abundant form of cold, observable gas in galaxies in the nearby
universe (e.g., \citealp{2009MNRAS.394.1857O}). HI observations
however can be time consuming, especially for galaxies with low
absolute gas mass.

Galaxies with low gas content can be of extremely different types:
gas-poor galaxies of all stellar masses and gas-rich galaxies with low
stellar masses. With existing flux-limited surveys such as the ALFALFA
21cm blind HI survey \citep{2011AJ....142..170H}, we cannot measure
the gas masses for these two populations beyond our nearest
neighbors. Fractional gas-mass limited surveys, such as the GALEX
Arecibo SDSS Survey (GASS; \citealp{2010MNRAS.403..683C}) and the
Nearby Field Galaxy Survey (NFGS;
\citealp{2010ApJ...708..841W,2013ApJ...777...42K}, hereafter K13),
allow us to examine galaxy gas content for a wider range of galaxy
types. Both of these data sets are representative of the galaxy
population in that they sample all types of galaxies within their
respective selection criteria. Neither of these two samples, though,
is a fair representation of the statistical distribution of galaxies
in the nearby universe. In contrast the RESOLVE (REsolved Spectroscopy
of a Local VolumE) survey is a complete volume-limited data set that
contains all galaxies above a ``cold baryonic'' (stellar + cold gas)
mass limit of \mbox{$\sim$10$^{9.1-9.3}$ \msun{}} (in two separate
subvolumes Kannappan et al.\ in prep.). The RESOLVE HI mass census is
also fractional mass limited (Stark et al.\ in prep.).

Already obtaining an HI mass census for the RESOLVE survey
($\sim$$1550$ galaxies) has required several hundreds of hours on
radio telescopes. To obtain gas masses for larger galaxy data sets, we
must develop accurate gas mass predictors. One particular use of such
estimators is to obtain galaxy cold baryonic masses, which are the
optimal indicator of dynamical mass for gas-rich galaxies (e.g., the
baryonic Tully-Fisher relation, \citealp{2000ApJ...533L..99M}). For
higher mass galaxies the baryonic component is dominated by the
stars. For lower mass galaxies, particularly below the gas-richness
threshold mass at \mbox{$\sim$10$^{9.7}$ \msun{}} in stellar mass,
galaxies can have as much cold gas as stars, or even be dominated by
their cold gas mass (K13).  It is important to characterize galaxy
mass, especially for dwarf galaxies, by cold baryonic mass (stars +
cold gas) rather than stellar mass alone.  For large imaging surveys,
such characterization will be impossible without the aid of accurate
gas mass predictors calibrated on existing galaxy surveys with
complete HI data.

%\subsection{Previous Gas Fractions}

One such predictor is the photometric gas fractions ``PGF'' technique,
which allows us to estimate galaxy cold gas mass primarily using
color. The PGF technique was first presented in
\citet{2004ApJ...611L..89K} as an observed relation between log
gas-to-stellar-mass ratio or G/S and \cuk{} color (see also
\citealp{2008AIPC.1035..163K}).  The relation between log(G/S) and
color is surprisingly tight: $\sigma \sim$ 0.37 dex.  This early work
on the PGF technique used a sample that cross-matched between a
flux-limited parent sample from imaging surveys and a heterogeneous
collection of available HI detections from the HyperLeda catalog
\citep{2003A&A...412...45P}. In \citet{2009MNRAS.397.1243Z}, the
authors used a similarly selected sample and find smaller scatter
$\sigma$ $\sim$ 0.3 dex using \cgr{} color and including $i$-band
surface brightness as a third parameter in the fit.

More recently, the GASS team has explored the PGF technique using
\cnuvr{} color combined with stellar mass surface density
\citep{2010MNRAS.403..683C} to create a ``gas fraction plane,''
finding $\sigma$ = 0.315 dex.  GASS is a stellar mass limited sample
that is representative of high mass galaxies and has measured HI
masses or upper limits down to a fixed fractional gas mass of
\mbox{1-5\%} of the stellar mass. Their PGF calibration, however, does
not accurately recover the HI masses for the bluest, most gas-rich
galaxies.  In \citet{2012A&A...544A..65C} and
\citet{2013MNRAS.436...34C}, the authors provide updated calibrations
excluding galaxies with \cnuvr{} $>$ 4.5, which yield smaller
residuals for gas-rich galaxies and smaller scatter overall $\sigma$ =
0.29 dex. To combat the residuals for gas-rich galaxies,
\citet{2012MNRAS.424.1471L} use the GASS sample to produced a
calibration from a combination of \cnuvr{} color, stellar mass,
stellar mass surface density, and \cgr{} color gradient. Their PGF
calibration more accurately predicts log(G/S) for gas-rich galaxies
from the flux-limited ALFALFA survey with $\sigma$ = 0.29 dex. The use
of multiple variables covariant with log(G/S) and each other, however,
prevents meaningful physical interpretation and artificially reduces
scatter.

The ALFALFA blind 21cm survey has also been used to derive a PGF
calibration by \citet{2012ApJ...756..113H}, who use S/N $>$ 6.5
reliable detections (code 1) and lower S/N detections with reliable
optical counterparts (code 2) from the $\alpha$.40 catalog
\citep{2011AJ....142..170H}. The calibration is based on \cnuvr{}
color and stellar mass surface density. Since the ALFALFA survey is
flux-limited, the calibration sample is biased towards gas-rich
objects and produces an offset towards higher gas fractions when
compared to the GASS PGF calibrations \citep{2012ApJ...756..113H}.

Lastly, K13 provides a PGF calibration for the Nearby Field Galaxy
Survey \citep{2000ApJS..126..271J}, a $B$-band selected,
representative galaxy survey that contains either HI detections or
strong upper limits for all galaxies.  The PGF calibration uses only
\cuj{} color and has scatter of $\sigma$ $\sim$ 0.34 dex. While the
scatter measured is higher than in other works, we note that the
calibration relies on color only and includes low-mass galaxies, which
have larger intrinsic uncertainties on their stellar mass estimates,
while GASS is limited to high stellar mass galaxies. K13 also shows
the effect of adding molecular gas for a subsample of the NFGS
galaxies, finding that for large spiral galaxies with low values of
log(G/S) the calibration is tightened when combining the molecular and
atomic hydrogen mass as the galaxy cold gas mass.

%Because the NFGS includes all galaxies from dwarfs to
%giants, it is not surprising that the scatter is higher than for
%calibrations relying on selection parameters that create more
%homogeneous data sets (e.g., high mass galaxies for GASS, gas-rich
%galaxies for ALFALFA). The NFGS calibration also relies on a single
%photometric parameter, \cuj{} color, making it more practical for
%estimating gas masses. 

The interpretation of the tight relation between color and log(G/S)
has been discussed in a few of these works. In
\citet{2004ApJ...611L..89K} the correlation between log(G/S) and
\cuk{} color is linked to the correlation between apparent $u$-band
magnitude and apparent HI magnitude. This correlation is understood as
the common link between the two quantities and the amount of recent
star formation within the galaxy.

% The PGF calibration in \citet{2004ApJ...611L..89K} is used
%to explore the gas content of galaxies near the transitional
%bimodality mass \citep{2003MNRAS.341...54K}, above which galaxies have
%quenched star formation and below which continuous star formation is
%more prevalent. For red sequence galaxies above the bimodality mass,
%gas fractions are very low (G/S $\sim$ 1:100), while below the
%bimodality mass gas fractions are more intermediate (G/S $\sim$ few
%times 1:10). For blue sequence galaxies above the bimodality mass, gas
%fractions are intermediate, while below the bimodality mass the gas
%content can equal to or even dominate the stellar mass content (G/S
%$\sim$ 1:1). Newer results from K13 reveal that the transition from
%intermediate gas fractions to gas-dominated galaxies actually occurs
%at a lower galaxy mass scale(M$_{star}$ = 10$^{9.7}$\msun, or the
%gas-richness threshold mass).

Another interpretation of the PGF relation comes from
\citet{2009MNRAS.397.1243Z}, who claim the PGF calibration is a
manifestation of the Kennicutt-Schmidt relationship between the
surface densities of star formation and of cold gas, which has been
calibrated on the short star formation timescales probed by H$\alpha$
\citep{1963ApJ...137..758S,1998ApJ...498..541K}. The results of K13,
however, show that the \cuj{} color of a galaxy can be interpreted
through stellar population modeling as the fractional stellar mass
growth rate, defined as the mass of stars formed in the last Gyr
divided by the pre-existing stellar mass. Thus \cuj{} color probes
timescales much longer than those probed by H$\alpha$.  In this light,
the \textit{current} galaxy gas reservoir is related to the galaxy's
\textit{past} growth rate over long timescales, and blue low-mass
galaxies, which typically have high gas-to-stellar mass ratios
(sometimes as much as 10), have been growing at rates inconsistent with
closed box models and requiring ongoing cosmic accretion (K13). The
authors argue that it is the long-term physics of accretion, rather
than the short-term physics of the Kennicutt-Schmidt relation, that
underlies the PGF technique.

In this work, we provide new z=0 PGF calibrations using the A-semester
of the volume-limited RESOLVE survey (RESOLVE-A). This data set offers
several key advantages over the previous calibrations discussed here.
First, we use newly reprocessed photometry, presented here, from
several imaging surveys. Superior photometry and well understood
systematic errors allows us to estimate reliable stellar masses
through SED fitting. Second, we have an almost complete (78\%) HI data
set for galaxies with detections or strong upper limits (defined here
as 1.4M$_{HI}$ $<$ 0.05M$_{star}$), and we are able to incorporate the
remaining 22\% that are confused or have weak upper limits through
statistical modeling using survival analysis. Third, our data set is
limited on absolute $r$-band magnitude, which most closely corresponds
to baryonic mass (K13), and the survey is complete to \mbox{M$_{bary}$
  $\sim$ 10$^{9.3}$ \msun}, well into the gas-dominated regime (see \S
\ref{sec:resolvea}). Lastly, because we use a volume-limited data set,
we correctly represent the number density of galaxies in the local
universe in color and log(G/S) parameter space.

% While the calibration is
%biased towards lower mass galaxies, it provides an accurate
%description of the galaxy population in the local universe.

%    A. volume-limited, baryonic-mass limited dataset 
%      1. M$_r$ $<$ -17.33 roughly correspoinding to $M_{bary}$ $<$ 10$^{9.3}$ \msun\
%      2. accurately represent dwarf statistics in a volume 

%    B. nearly complete HI dataset
%      1. 85\% complete HI, some percentage hopelessly confused, some hardest reddest dwarfs - use a scheme to include them.

%\subsection{Roadmap}

This paper is organized as follows.  First we describe the RESOLVE
survey and its two subvolumes in \S \ref{sec:datasets}.  Next we
detail the reprocessed photometry, stellar mass estimation, and HI
data in \S \ref{sec:data}.  We then describe color-limited PGF
calibrations using linear fits in \S \ref{sec:simplecals} and examine
correlations between log(G/S) residuals and photometric parameters to
obtain tighter calibrations in \S \ref{sec:3pcorr}. The linear fits
are limited by their inability to predict gas masses for red galaxies,
for which the correlation between color and log(G/S) breaks down, as
well as by the fact that we cannot simply include galaxies with weak
upper limits. To properly predict gas masses for all galaxies, we
describe in \S \ref{sec:cals} a new PGF calibration using a 2D model
to fit to a density field, yielding log(G/S) probability distributions
for galaxies of all colors.  In \S \ref{sec:discussion} we test the
new calibrations on the RESOLVE-B data set and we compare with
previous calibrations from the literature, finding that our new
calibrations are not biased for a z=0, volume-limited survey. Lastly
we summarize our main conclusions in \S \ref{sec:conclusions}.

%A. present RESOLVE sample

%    B. present photometric, stellar mass and HI data for calibration

%    C. present methodology, examination of 3rd parameters, and new calibrations

%    D. compare with previous calibrations from the literature

%    E. discussion and conclusions

%%%%%%%%%%%%%%%%%%%%%%%%%%%%%%DATA SETS%%%%%%%%%%%%%%%%%%%%%%%%%%%%%
\section{Data Sets}
\label{sec:datasets}

For this work, we use the RESOLVE survey
(\citealp{2008AIPC.1035..163K}; Kannappan et al.\ in prep.), a
volume-limited mass census, to create and test new PGF
calibrations. The RESOLVE survey is ideal for calibrating gas mass
estimators because it has a complete galaxy census with nearly
complete HI data down to fixed fractional mass limits.

\begin{figure*}
\epsscale{1.0}
%\plottwo{/srv/one/keckert/papers/govers/figs3/fig1a.eps}{/srv/one/keckert/papers/govers/figs3/fig1b.eps}
\plottwo{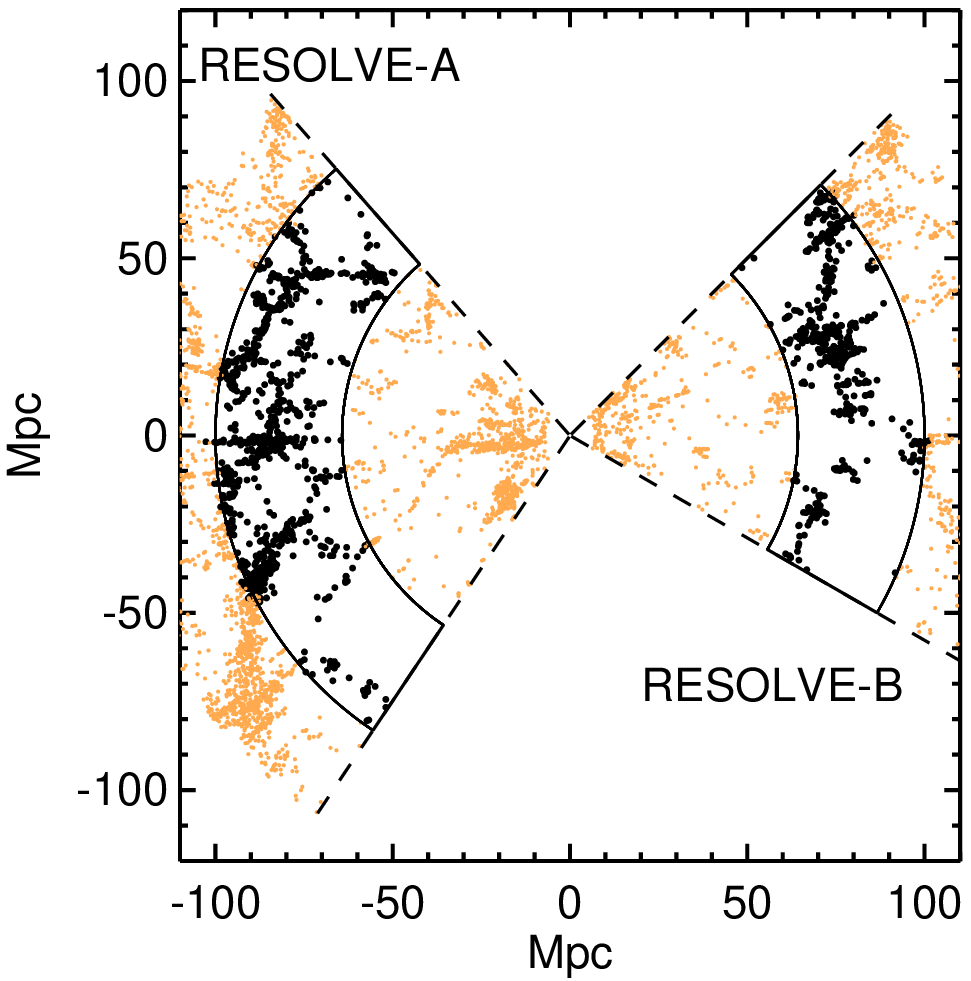}{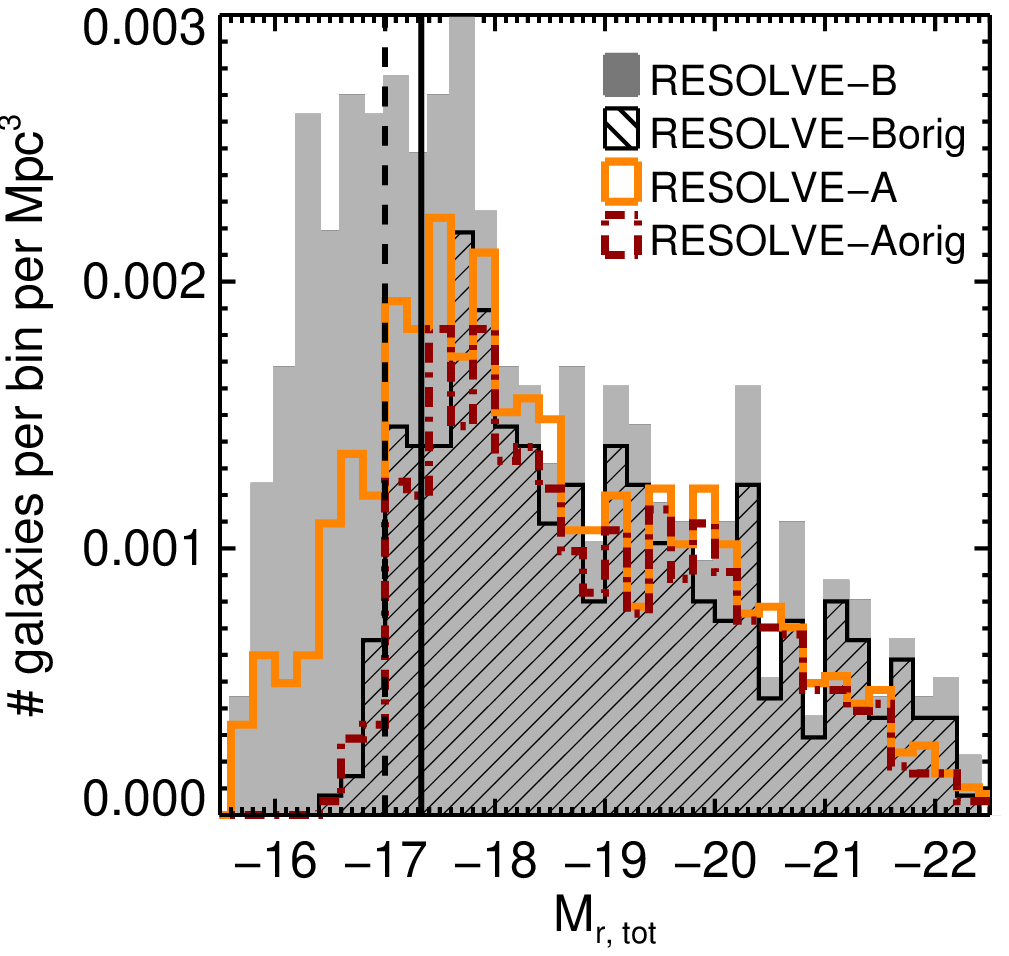}

\caption{RA--cz and $r$-band absolute magnitude distributions of
  RESOLVE-A and -B semesters.  a) The regions outlined in black show
  the RA and cz limits of the two RESOLVE subvolumes; both have been
  collapsed in Dec, ranging 0--5$^{\circ}$ for RESOLVE-A and $-1.25$ to
  +1.25$^{\circ}$ for RESOLVE-B.  The black dots show RESOLVE
  galaxies, identified as members of RESOLVE because their group
  redshift falls within the limits of the survey. Orange points show
  galaxies belonging to groups outside of the RESOLVE cz limits
  4500--7000 \kms. b) The black hash filled histogram and red
  dot-dashed outlined histograms show the original absolute $r$-band
  magnitude distributions for RESOLVE-B and RESOLVE-A respectively.
  Both distributions fall off rapidly below M$_{r, tot}$ $<$ $-17.33$.
  The grey shaded histogram and the orange solid outlined histogram
  show the full absolute $r$-band magnitude distributions for
  RESOLVE-B and RESOLVE-A after redshift completion efforts described
  in \S \ref{sec:datasets}.  The RESOLVE-A region is still complete
  only to M$_{r,tot}$ = $-17.33$, however we are able to move the
  RESOLVE-B completeness limit down to $-17.0$. }

\label{fg:resolvebutterfly}
\end{figure*}

RESOLVE is an equatorial survey covering two semesters (RESOLVE-A and
RESOLVE-B) shown in Figure \ref{fg:resolvebutterfly}a. The RESOLVE
survey is located within the SDSS footprint and makes use of the SDSS
redshift survey to build up survey membership with completeness down
to \mbox{M$_{r, petro}$ = $-17.23$}, the absolute $r$-band magnitude
corresponding to the SDSS survey limit of \mbox{m$_{r, petro}$ =
  17.77} at the outer RESOLVE cz limit, \mbox{7000 \kms}. We also
include additional redshifts from various archival sources: the
Updated Zwicky Catalog (UZC, \citealp{1999PASP..111..438F}), HyperLeda
\citep{2003A&A...412...45P}, 6dF \citep{2009MNRAS.399..683J}, 2dF
\citep{2001MNRAS.328.1039C}, GAMA \citep{2011MNRAS.413..971D},
ALFALFA \citep{2011AJ....142..170H}, and RESOLVE
observations (Kannappan et al.\ in prep.). These extra redshifts
provide greater completeness to the RESOLVE data set, as detailed in a
companion paper on the baryonic mass function and its dependence on
environment (Eckert et al.\ in prep.) and in the RESOLVE survey design
paper (Kannappan et al.\ in prep.).  For both RESOLVE-A and RESOLVE-B
we have custom reprocessed photometry providing total magnitudes and
systematic errors for \textit{GALEX} NUV (plus new
  \textit{Swift} UVOT imaging for nineteen galaxies), SDSS $ugriz$,
UKIDSS $YHK$, and 2MASS $JHK$ bands as available (described in \S
\ref{sec:photdata}).

To define survey membership, we use the redshift of the group to which
each galaxy belongs.  Group finding is performed using the
Friends-of-Friends algorithm from \citet{2006ApJS..167....1B} with on
sky and line of sight linking lengths of 0.07 and 1.1 respectively as
suggested by \citet{2014MNRAS.440.1763D} and also justified in Eckert
et al.\ (in prep.).  As can be seen in Figure
\ref{fg:resolvebutterfly}a, galaxies with redshifts nominally outside
the volume may be grouped with galaxies inside the volume, while
occasionally galaxies with nominal redshifts inside the volume may be
removed as they belong to a group outside the volume.

The gas data used in this paper come from the RESOLVE HI census, which
is described in \S \ref{sec:himasses} and will be published in Stark
et al.\ (in prep.). RESOLVE HI observations build on the ALFALFA blind
21cm survey \citep{2011AJ....142..170H}, which covers the entire
RESOLVE-A footprint and partially covers the RESOLVE-B footprint. New
pointed observations with the GBT and Arecibo telescopes follow up on
marginal detections, sources with weak upper limits, or sources with
no HI data.

%    A. The RESOLVE survey covers 2 semesters \citealp{2008AIPC.1035..163K}; Kannappan et al, in prep)
%       1. volume-limited
%       2. dwarf-dominated
%       3. mass-census - paper represents stars and gas, eventually dynamical mass

\subsection{RESOLVE-A}
\label{sec:resolvea}

The RESOLVE-A data set shown in Figure \ref{fg:resolvebutterfly}a
occupies a volume of $\sim$38,400~Mpc$^3$ defined by:
\mbox{131.25$^{\circ}$ $<$ RA $<$ 236.25$^{\circ}$}, 0$^{\circ}$ $<$
Dec $<$ 5$^{\circ}$, and \mbox{4500 \kms{} $<$ cz $<$ 7000
  \kms{}}. The data set's $r$-band absolute magnitude distribution is
shown in the orange solid line histogram in Figure
\ref{fg:resolvebutterfly}b.  RESOLVE-A is complete down to M$_{r,tot}$
$<$ $-17.33$ using the reprocessed photometry described in \S
\ref{sec:photdata}. A magnitude of \mbox{M$_{r,tot}$ $\sim$ $-17.33$}
roughly corresponds to M$_{bary}$ $\sim$ 10$^{9.1}$ \msun{} (K13). To
determine the baryonic mass completeness limit, we consider the
scatter in baryonic mass-to-light ratio, which can be at least as high
as 3 resulting in a baryonic mass limit of 10$^{9.3}$~\msun.  The
RESOLVE-A survey contains 955 galaxies brighter than
M$_{r, tot}$ = $-17.33$. Of these 955 galaxies,
$\sim$12\% were added from redshift surveys besides
the SDSS main redshift survey.  The data set resulting from the SDSS
main redshift survey alone (RESOLVE-Aorig) is shown as the red
dot-dashed line histogram in Figure \ref{fg:resolvebutterfly}b. The
RESOLVE-A region is 78\% complete in HI when counting
successful, unconfused HI detections and strong upper limits resulting
in 1.4M$_{HI}$ $<$ 0.05M$_{star}$.  We use the RESOLVE-A data set to
determine our PGF calibrations, accounting for missing HI data with an
iterative Monte Carlo technique akin to survival analysis (see \S
\ref{sec:simplecals} and \S \ref{sec:cals}).

\subsection{RESOLVE-B}
\label{sec:resolveb}

The RESOLVE-B data set is located in the SDSS Stripe 82 equatorial
region, and it occupies a smaller volume of $\sim$13,700 Mpc$^3$
defined by: 22h $<$ RA $<$ 3h, $-1.25^{\circ}$ $<$ Dec $<$
1.25$^{\circ}$, and \mbox{4500 \kms{} $<$ cz $<$ 7000 \kms{}}.  In
Figure \ref{fg:resolvebutterfly}b the absolute $r$-band magnitude
distribution is shown for RESOLVE-B galaxies coming from the SDSS main
redshift survey as the black hashed histogram (RESOLVE-Borig), as well
as for the full RESOLVE-B data set (grey filled histogram), which
includes redshifts from the sources mentioned in \S \ref{sec:datasets}
and extra SDSS redshift observations over the Stripe 82 footprint. The
data set is complete in $r$-band absolute magnitude down to
\mbox{M$_{r, tot}$ $\cong$ $-17.0$}, slightly deeper than RESOLVE-A
implying completeness to \mbox{M$_{bary}$ $\sim$ 10$^{9.1}$
  \msun}. The RESOLVE-B survey contains 487 galaxies
to this limit, $\sim$25\% of which have been added by
redshift surveys besides the SDSS main redshift survey. We have
recovered more galaxies in RESOLVE-B than in RESOLVE-A due to the
extra spectroscopic passes done by the SDSS that are not part of the
main SDSS redshift survey. The RESOLVE-B region is
$\sim$75\% complete in HI data for good HI detections
and strong upper limits.  We use the RESOLVE-B data set to test our
new PGF calibrations and compare with other calibrations from the
literature (see \S \ref{sec:discussion}).

\section{Data}
\label{sec:data}

For this work we need consistent and well calibrated photometry,
stellar masses, and HI masses down to fixed fractional mass limits.
We present our methods for reprocessing UV, optical, and IR photometry
for the RESOLVE survey in \S \ref{sec:photdata}.  We then describe our
stellar mass estimation through SED modeling in \S \ref{sec:mstar}.
Lastly we describe the various sources of HI data and the measurement
of HI masses \S \ref{sec:himasses}.

\subsection{Photometric Data}
\label{sec:photdata}

We have reprocessed photometric data for the RESOLVE survey from the
UV to near IR to obtain consistent, well-determined total magnitudes,
and we use two to three methods of flux extrapolation per band to
characterize systematic errors on the total magnitudes of the
galaxies. We have also run the same pipeline on the larger
volume-limited ECO (Environmental COntext) catalog (Moffett et al.,
submitted), which surrounds the RESOLVE-A subvolume. We use optical
$ugriz$ data from SDSS \citep{2011ApJS..193...29A}, NIR $JHK$ from
2MASS \citep{2006AJ....131.1163S} and/or $YHK$ from UKIDSS
\citep{2008MNRAS.384..637H}, and NUV from the \textit{GALEX} mission
\citep{2007ApJS..173..682M}. Our NUV data are mostly MIS depth due to
prioritization of the RESOLVE-A footprint late in the \textit{GALEX}
mission (after the FUV detector failed), while RESOLVE-B (Stripe 82)
already had deep coverage in both the NUV and FUV for other
programs. The SDSS optical imaging in the RESOLVE-B footprint is extra
deep due to repeated imaging with typically 20 frames per location on
the sky \citep{2014ApJ...794..120A}. With our improved photometry and
realistic error measurements, we are able to measure reliable colors
and perform accurate stellar mass estimation via SED modeling.

Our reprocessed photometry improves over SDSS pipeline photometry in
several key ways.  First, we use images with improved sky subtraction
coming from either \citet{2011AJ....142...31B} for SDSS or our own
additional sky subtraction for 2MASS and UKIDSS. Second, we use the
sum of the high S/N $gri$ images to define the elliptical apertures,
allowing us to determine the PA and axial ratio of the outer disk if
present. Third, we apply these same elliptical apertures to all bands
which allows us to measure magnitudes for galaxies that may not have
been detected by the original automated survey pipeline in certain
bands, especially low surface brightness galaxies in 2MASS, UKIDSS,
and \textit{GALEX}.  Lastly, we use two to three non-parametric
methods of total magnitude extrapolation, measuring the light from
each band independently (see Figure \ref{fg:threemethods} and \S
\ref{sec:magextr}). This last point allows for color gradients within
galaxies, as opposed to the model magnitudes provided by SDSS (more
details in \S \ref{sec:compsdss}), and allows us to measure systematic
errors on magnitudes.

We provide a comparison of the magnitudes, colors, and radii with
photometry from the DR7 catalog of SDSS in Figure
\ref{fg:photcompmagrad} and \S \ref{sec:compsdss}. Briefly
summarizing, we find that the newly reprocessed photometry yields
brighter magnitudes and larger effective radii.  The colors tend to be
bluer for large objects, which we believe to be a consequence of both
the improved sky subtraction from \citet{2011AJ....142...31B} and the
fact that we allow color gradients in magnitude estimation.  The newly
reprocessed photometry does not create a tight red sequence on the
color-magnitude relation as seen in the DR7 photometry, however we
argue that the tight red sequence may be a consequence of these two
issues in \S \ref{sec:compsdss}. We also discuss an independent
validation of our methods with the NFGS survey (shown in Figure 2a of
K13) in \S \ref{sec:compsdss}.

In addition to the reprocessed photometry, this paper also presents
new UV observations of 19 galaxies using the \textit{Swift}
Ultraviolet/Optical Telescope (UVOT, \citealp{2005SSRv..120...95R},
see also \citealp{2004ApJ...611.1005G}).  We use imaging from the uvm2
filter, which has a comparable central wavelength but narrower width
than the \textit{GALEX} NUV filter (see
\citealp{2008MNRAS.383..627P}). Compared to \textit{GALEX}, the
pointing restrictions for UVOT are much less stringent, allowing us to
obtain observations during \textit{Swift} team fill-in time for
RESOLVE-B galaxies that were not observed by \textit{GALEX} or had
only AIS depth ($\sim$150s) coverage. Nineteen galaxies were observed
for more than 1 ks, the minimum exposure for useful photometry. Images
were processed following \citet{2011AJ....141..205H}. Each galaxy was
manually inspected to make sure that the surface brightness in the
uvm2 band was low enough that the resulting photometry errors due to
coincidence loss were below 1\% (see
\citealp{2008MNRAS.383..627P,2010MNRAS.406.1687B,2011AJ....141..205H}). We
apply similar photometric processing to the \textit{Swift} data as to
the archival data including matched elliptical apertures and multiple
extrapolation techniques.

\subsubsection{Custom Processed Data}
\label{sec:datareproc}

We start the photometric reprocessing by downloading the data from
each respective website, performing background subtraction and
coaddition when necessary, and cropping a region around the galaxy 9
times the Petrosian 90\% light radius as reported by SDSS with a
minimum crop size of 3x3 arcmin$^{2}$. Because some galaxies are quite
large, we rescale images to a fixed image size, causing the pixel
scale to vary from galaxy to galaxy.  Since the Stripe 82 region has
been repeatedly observed in $ugriz$, for RESOLVE-B galaxies we coadd
the many frames of data by inversely weighting by the variance of sky
fluctuations using the IRAF task \textit{imcombine}.  Within the
RESOLVE-A region there is typically one $ugriz$ frame per region of
sky, and we use SWARP \citep{2002ASPC..281..228B} to stitch together
adjacent frames when necessary, averaging together pixels where there
is overlap between images. No additional background subtraction is
done, as we are using SDSS DR8 images with the optimized sky
background subtraction of \citet{2011AJ....142...31B}. For 2MASS $JHK$
and UKIDSS $YJHK$, we perform additional background subtraction by
fitting and subtracting a 3rd order polynomial to a region of the
galaxy frame where the galaxy and other objects are masked
out. Coaddition is similar to the single frame SDSS process, using
SWARP to stitch together 2MASS and UKIDSS frames with a simple average
to combine pixels in overlapping areas of the sky. Based on visual
inspection of the UKIDSS data, we do not use the $J$ band due to
background subtraction and other issues that affect $\sim$75\% of the
data. We have also examined the $YHK$ images for each galaxy by eye to
flag any cases with bad data. The \textit{GALEX} NUV images do not
require additional background subtraction, and these images are simply
coadded using SWARP and weighted by exposure time. For the
\textit{Swift} uvm2 images we use Source Extractor to identify and
mask objects within the frame, then subtract off the median level of
the non-masked areas as the sky-background.

A significant number of galaxies ($\sim$16\%) in RESOLVE have
half-light diameters smaller than three times the typical $r$-band psf
FWHM of $\sim$$1.4$\arcsec, warranting psf-matching across the optical
bands and UKIDSS IR bands. For each galaxy, we use the SDSS provided
psField frame to reconstruct the psf for each band at the galaxy
position on the frame. First we identify the band with the worst psf
seeing (typically $u$ or $g$). Next we find the Gaussian $\sigma$
value with which to convolve the psf of each given band to eliminate
the difference between the psf of the worst band and that given band.
This Gaussian $\sigma$ value is then used to create a Gaussian kernel
that is convolved with the galaxy cropped image. Since we have changed
the pixel scale of the frames of larger galaxies, we make sure to
convert the $\sigma$ value into the correct pixel scale for that
galaxy. For UKIDSS IR data a similar procedure is run, except that the
psf for each frame is constructed from stars identified by Source
Extractor. If the value of the converted $\sigma$ value is less than
one pixel, we do not perform the convolution. We do not psf-match the
NUV, uvm2, or 2MASS $JHK$ bands because their typical psfs are much
larger than the SDSS ($\sim$5.5\arcsec{} for NUV, $\sim$2.5\arcsec{}
for uvm2, and $\sim$2\arcsec{} for 2MASS). Thus aperture matched
magnitude measurements for the UV and 2MASS IR will not be correct,
especially for small galaxies.

Masks are made from the $r$-band image using Source Extractor
\citep{1996A&AS..117..393B} to detect stars and galaxies other than
the target. Masks are checked by eye to ensure there is no under/over
masking. In \S \ref{sec:compsdss} we discuss the application of this
photometric reprocessing pipeline for the ECO (Environmental COntext)
catalog (Moffett et al., submitted), for which we do not check each
mask by eye. Instead we use a first iteration of the pipeline to check
for discrepant total and aperture magnitudes as well as cases with no
valid magnitudes at the end. For the most egregious outliers, we check
the masks and edit by hand where necessary. 

%The comparison of totalM$_{r,tot}$ is shown in Figure \ref{fg:compareweco}.

To determine the parameters of the elliptical fit (namely the PA and
ellipticity), we use an iterative procedure involving two programs.
First, Source Extractor is run on each galaxy's cropped $r$-band frame
to find an initial guess for the center, PA, and ellipticity, and 90\%
light radius.  Second, we run the IRAF task \textit{ellipse} on the
$gri$ coadded image using the Source Extractor quantities as inputs
and allowing the PA and ellipticity, but not the center, to vary. The
$gri$ images have the highest signal-to-noise data, and by coadding
these three bands, we provide the best image to feed to
\textit{ellipse} for determining the PA and ellipticity in the the
outer parts of the galaxy. The final PA and ellipticity are chosen
using a median of the fits from the outer disk of the galaxy, where
``outer disk'' is defined between one and two times the 90\% $r$-band
light radius determined by Source Extractor. Using the final PA,
ellipticity and galaxy center, we then perform a fixed
\textit{ellipse} fit on the $gri$ summed image to determine the set of
annuli over which to measure the galaxy surface brightness profiles
for each band separately.

These same annuli are then automatically applied to the
\textit{GALEX}, SDSS, 2MASS, and UKIDSS data.  To match the NUV and
$YJHK$ images to the SDSS images, we resample them to the pixel scale
of the SDSS image.  Imposing the same annuli over all bands allows us
to measure the galaxy light out to its furthest extent (based on
$gri$).  For the IR bands, we are able to measure magnitudes for twice
as many galaxies as the 2MASS catalog detects and for $\sim$1.15 times
as many galaxies as the UKIDSS catalog (based on public DR8plus).

Some galaxies are in very close pairs or embedded within a larger
galaxy. To obtain better photometry for these galaxies, we have
attempted to subtract off the galaxy light from the interfering galaxy
for each frame, when it seems possible to identify the light belonging
to a specific galaxy. To flag such cases we search for whether a
nearby galaxy is within 4 times the 50\% $r$-band light radius of each
RESOLVE galaxy. This selection returns 27 systems. We inspect these 27
systems and remove 11 which appear well separated from their
neighbors. We also remove three closely paired systems (rs1158/rs1160,
rs0100/rs0101, rs0196/rs0197), two that involve merging star forming
spirals and one that contains two similarly sized elliptical galaxies,
all three of which are so close as to make it impossible to
disentangle the light from each galaxy. We find 13 systems that
benefit from attempting to remove the light from a galaxy.

%rs0222 - separated
%rs0436/rs0437 - separated
%rs1333 - separated
%rs0670 - separated
%rs0800 - separated
%rs1197 - separated
%rf0011- separated
%rf0113- separated
%rf0312 - separated
%rf0346 - separated
%rf0355/rf0357 - separated
%rs1160 - messy
%rs0100 - messy
%rs0196/rs0197 - too overlapping
%(rs1320)rs0072/rsnew9 - embedded
%rs0639 - embedded
%rs0673/rs0675 - embedded
%rs1089/rs1090 - embedded
%rs1226/rs1227 - embedded
%rs1232/rs1233 - embedded
%rf0090/rf0094 - embedded
%rs0267/rs0268 - pair
%rs0397/rs0398 - pair
%rs0749/rs0750 - pair
%rs0851/rs0852 - pair
%rf0015/rf0016 - pair
%rf0309 - pair

\textit{Embedded Galaxies:} There are five systems that are heavily
embedded inside a much larger galaxy: rs0675, rs0749, rs1233, rs1227, and
rs0072 (inside rs0673, rs0750, rs1232, rs1226, and rs0072
respectively). Another three systems are on the outskirts of a larger
galaxy: rs0639, rs1089, and rf0090 (just outside of rs0642, rs1090,
and rf0094 respectively). To obtain better photometry for these eight
embedded galaxies, we first mask the small embedded galaxy and run
\textit{ellipse} on the larger galaxy. We then subtract off the model
flux from the larger galaxy that is output from \textit{ellipse}. We
use the resulting image that has the large galaxy subtracted out to
run through the procedures described in this section, ensuring that
any residuals from the model are masked out. 

\textit{Close Pairs:} There are five close pair systems for which
subtracting off the light of one or both of the members improves
the magnitude estimates. These systems are rs0267/rs0268, rs0397/rs0398,
rs0851/rs0852, rf0015/rf0016, and rf0309/rf0310. To subtract off the
light of each member we use the following steps:

\textbullet{} First, we identify the galaxy with the simpler light
profile, which we call galaxy-A. For each pair galaxy-A is rs0268,
rs0397, rs0851, rf0016, and rf0310.

\textbullet{} Second, we start with the image for the other
galaxy, which we call galaxy-B. We mask galaxy-B and run
\textit{ellipse} to fit the light profile of galaxy-A.

\textbullet{} Third, we subtract off the galaxy-A model flux as
provided by \textit{ellipse} from the galaxy-B image. The resulting
image is used in the standard pipeline for galaxy-B (rs0267,
rs0398, rs0852, rf0015, rf0309).

\textbullet{} In most cases, we do not then subtract off the galaxy-B
image for galaxy-A. We choose not to for a variety of reasons. For
rs0267/rs0268, galaxy-B (rs0267) does not have a regular light profile
making it difficult to subtract off. For rs0397/rs0398 and
rf0309/rf0310 galaxy-B (rs0398, rf0310) is edge-on and easy to
mask. For rs0851/rs0852, galaxy-B (rs0852) is much smaller and easier
to mask out.

\textbullet{} For the last pair rf0015/rf0016, we take the galaxy-B
(rf0015) image with galaxy-A's light subtracted off and mask any
residuals from galaxy-A. We run \textit{ellipse} to fit the light
profile of galaxy-B.  We then subtract off the model fit to galaxy-B
provided by \textit{ellipse} from the original galaxy-B image. The
resulting image of galaxy-A (which is no longer in the center) is
run through the pipeline with newly generated masks.

We perform these procedures for all optical bands and UKIDSS
images. For 2MASS and \textit{GALEX}, we check first whether the
subtraction is needed because the galaxy light may not extend far
enough in these bands.

\subsubsection{Magnitude Extrapolation}
\label{sec:magextr}

\begin{figure*}
\epsscale{1.}
%\plotone{/srv/one/keckert/papers/govers/figs3/fig2.eps}
\plotone{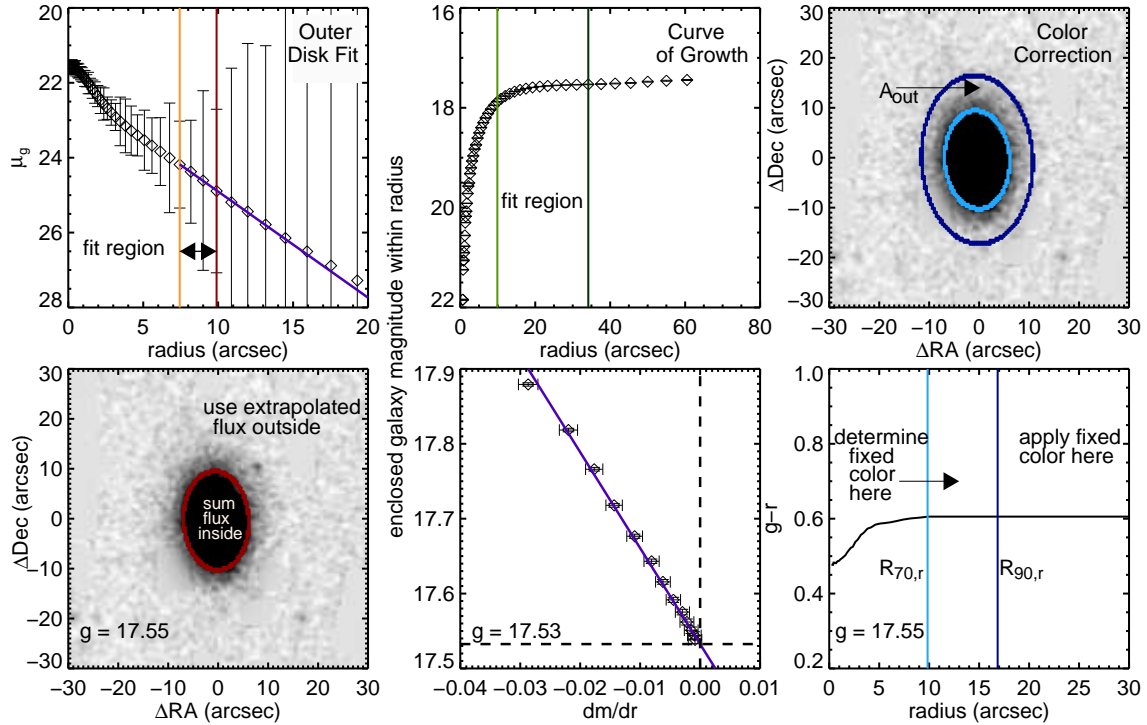}
\caption{Illustration of the three methods described in \S
  \ref{sec:magextr} to extrapolate the total $g$-band magnitude for
  RESOLVE-B galaxy rf0218. In this case, all three estimates agree
  closely (17.55, 17.53, 17.55), yielding a small systematic error
  (0.021).  Left column: To demonstrate the Outer Disk Fit method, we
  show in the top left panel annular $g$-band surface brightness
  vs.\ radius with the fitting region marked by the orange (inner) and
  red (outer) lines.  The blue line shows the exponential disk fit to
  the data points.  The bottom left panel illustrates how we compute
  the total magnitude as the sum of the raw galaxy flux inside the
  radius marked by the red line (same as radius in top panel marked by
  red line) and the extrapolated flux outside that radius.  Middle
  column: To demonstrate the Curve of Growth method, we show in the
  top middle panel enclosed galaxy magnitude vs.\ radius with the
  fitting region marked by light green (inner) and dark green (outer)
  lines.  The lower middle panel shows the enclosed galaxy magnitude
  vs.\ the derivative of enclosed magnitude with respect to radius for
  the points within the fitting region defined above.  The blue line
  shows the fit and the y-intercept at dm/dr = 0 is the total galaxy
  magnitude. Right column: To demonstrate the Outer Disk Color
  Correction method, we show in the top right panel the annulus
  A$_{out}$ defined by the $r$-band 70\% and 90\% light radii (light
  and dark blue lines respectively).  Using the flux ratio in
  A$_{out}$, we fix the color of the galaxy to determine the $g$-band
  flux beyond $R_{90,r}$, from the $r$-band flux in that region.}
\label{fg:threemethods}
\end{figure*}

To extrapolate total magnitudes from the fixed ellipse fits for the
optical bands, we use three methods: an exponential (Sersic index
$n=1$) fit to the outer disk, a non-parametric Curve of Growth
extrapolation, and an Outer Disk Color Correction based on the
$r$-band.  Figure \ref{fg:threemethods} shows schematics of all three
methods in the $g$ band for a RESOLVE-B galaxy.

\textit{Outer Disk Fit:} To compute the outer disk flux, we first
define a fitting region where the annular flux is 1 to 5 times the
$\sigma$ of the sky noise.  If the galaxy frame has been masked
heavily (due to nearby stars or galaxies), we use a region 3 to 8
times the $\sigma$ of the sky noise, and in extreme cases where a
bright star or galaxy is on top of the galaxy, we use a region defined
by 20-50 times the $\sigma$ of the sky noise.  Then we fit an
exponential disk function to the fitting region of the galaxy surface
brightness profile (between orange and red lines in left panel of
Figure \ref{fg:threemethods}) and sum the extrapolated flux from the
inner edge of the last ellipse in the fitting region (red line) to an
extremely large radius ``$R_{\infty}$'' or 1000\arcsec{} past the
inner edge of the last ellipse in the fitting region (red line).  To
compute the inner disk flux, we sum the raw, unmodeled flux interior
to the inner edge of the last ellipse in the fitting region (red
ellipse in left panels of Figure \ref{fg:threemethods}).  If pixels
are masked in the raw data, we replace those values with the model
output from \textit{ellipse}.  The exponential total magnitude equals
the sum of the measured inner flux and the extrapolated outer disk
flux.  The typical dividing radius is near the 90\% light radius in
the $r$ band.

\textit{Curve of Growth:} The Curve of Growth method, following
\citet{2009ApJ...703.1569M}, computes the total magnitude using the
derivative of the enclosed magnitude as a function of the radius.  The
enclosed magnitudes are calculated from the \textit{ellipse} profile.
A line is then fitted to the derivative of enclosed magnitude with
respect to the radius vs.\ the total enclosed magnitude. The line is
fitted over a new fitting region that extends farther out in the
galaxy profile, to where changes in the total enclosed magnitude as a
function or radius are small (between the light and dark green lines
in the central panel of Figure \ref{fg:threemethods}).  The
y-intercept of the fitted line, where dm/dr = 0, is the total Curve of
Growth magnitude.

\textit{Outer Disk Color Correction:} This method scales the outer
disk $r$-band flux to determine the outer disk flux of an object in
another band.  First, we use either the Curve of Growth or the Outer
Disk Fit $r$-band magnitude to determine the radii containing 70\% and
90\% of the $r$-band light in the running total flux profile (light
and dark blue ellipses show these respective radii in right most panel of Figure
\ref{fg:threemethods}).  If the galaxy frame is heavily masked (more
than 5\% of the image pixels), we prefer the $r$-band exponential
magnitude, otherwise the $r$-band Curve-of-Growth magnitude is used.
We next measure the galaxy flux within the $R_{70,r}$ to $R_{90,r}$
annulus (A$_{out}$).  If the S/N of the flux in this annulus is not
greater than 10, we decrease the inner radius of A$_{out}$ by
increments of 5\% down to 50\% of the $r$-band light, stopping when we
achieve S/N $>$ 10.  If the S/N is still less than 10 between
$R_{50,r}$ and $R_{90,r}$, we do not compute the galaxy magnitude with
this method.  Otherwise, we calculate the flux ratio between a given
band $x$ and the $r$ band within the annulus A$_{out}$, and we assume
that this ratio continues out to infinity.  From the $r$-band flux
from $R_{90,r}$ to $R_{\infty}$, and the flux ratio within A$_{out}$,
we estimate the flux in band $x$ from $R_{90,r}$ to $R_{\infty}$, then
add this flux to the raw enclosed flux inside $R_{90,r}$ to get the
final magnitude.

Extrapolation of the NIR and UV magnitudes proceeds similarly to the
optical extrapolation, but with a few subtleties. The Curve of Growth
method is the preferred method for our NIR data due to the poor
signal-noise for low surface brightness galaxies. Exponential fits are
used to determine whether or not the Curve of Growth method works
well.  If the two fits disagree significantly or if the object's
magnitude is very faint, we look at the magnitude calculated based on
the $i$ band (using the Outer Disk Color Correction method).  If
either the Curve of Growth or exponential matches the aperture
magnitude, that is chosen.  If neither method agrees, the Outer Disk
Color Correction is used and given a large systematic error ($>$0.5
mag). For the UV data from both \textit{GALEX} and
  \textit{Swift}, we find that the Curve of Growth method is the most
reliable magnitude estimation method as the clumpiness of the UV and
the possibility of XUV disks (extended UV emission outside the typical
optical extent of the galaxy; \citealp{2007ApJS..173..538T}) make
exponential disk fitting and fixing the outer disk color
impractical. The Outer Disk Color Correction method is also hampered
by the mismatch in psf between the UV images ($\sim$5.5\arcsec{}
and $\sim$2.5\arcsec{} for the NUV and uvm2
  respectively) and the psf of the convolved SDSS and UKIDSS images
($\sim$1.8\arcsec).  If the Curve of Growth method fails, though, we
use the magnitude of the Outer Disk Color Correction method, with a
systematic error $>$ 0.06 applied.

Errors for all bands are computed using not only the formal
statistical error on the magnitude, but also the systematic error
based on the difference in flux measured from the three methods. We
apply a built in floor for the systematic error based on the overall
distribution of systematic errors for the galaxy data set, such that
none are lower than the original 25 percentile.

In addition we compute half light and 90\% light radii in the $r$ band
($R_{50,r}$ and $R_{90,r}$), as well as the $r$-band surface
brightness within these radii ($\mu_{r,50}$ and $\mu_{r,90}$). We also
measure aperture magnitudes for all available bands within the
$r$-band half light and 90\% light radii, although the lack of psf
correction for the 2MASS $JHK$ and NUV and uvm2 bands
compromises associated aperture matched colors. We also compute the
$g-r$ color gradient (hereafter $\Delta_{g-r}$), which is defined as
the $g-r$ color within the annulus between the half light and 75\%
$r$-band light radii minus the $g-r$ color within the $r$-band half
light radius.  More positive colors indicate galaxies with bluer
centers.

Throughout this work we use Milky Way foreground extinction
corrections determined from the dust maps of
\citet{1998ApJ...500..525S} with the extinction curves of
\citet{1994ApJ...422..158O} for the optical and IR data, and of
\citet{1989ApJ...345..245C} for the NUV and uvm2
  data. For the NUV and uvm2 data we use the
extinction correction calculated at 2271 \AA{} and
  2221 \AA, the effective wavelengths of the NUV and uvm2 filter
  respectively. We note that using the more recently computed
extinction coefficients from \citet{2011ApJ...737..103S}, which use
the extinction curve from \citet{1999PASP..111...63F}, yields colors
that are $\sim$0.015 mag bluer in \cur{} ($\sim$0.04 bluer in \cnuvr)
and do not change the stellar mass estimates from \S \ref{sec:mstar}.

Table \ref{tb:phottable} provides descriptions of the columns that are
provided in a machine readable table with the photometry for the
RESOLVE survey. All galaxies processed are provided, including those
in the buffer and fainter than the nominal RESOLVE limits.

\begin{deluxetable}{ll}
\tablecaption{RESOLVE Custom Photometry Catalog Description}
\tablehead{\colhead{Column} & \colhead{Description}}
\startdata
1 & RESOLVE ID \\
2 & Right Ascension \\
3 & Declination \\
4 & cz \\
5 & group cz \\
6 & absolute SDSS $r$-band magnitude \\
7 & apparent SDSS $u$-band magnitude \\
8 & apparent SDSS $u$-band magnitude error \\
9 & apparent SDSS $g$-band magnitude \\
10 & apparent SDSS $g$-band magnitude error \\
11 & apparent SDSS $r$-band magnitude \\
12 & apparent SDSS $r$-band magnitude error \\
13 & apparent SDSS $i$-band magnitude \\
14 & apparent SDSS $i$-band magnitude error \\
15 & apparent SDSS $z$-band magnitude \\
16 & apparent SDSS $z$-band magnitude error \\
17 & apparent \textit{GALEX} NUV-band magnitude \\
18 & apparent \textit{GALEX} NUV-band magnitude error \\
19 & apparent \textit{Swift} uvm2-band magnitude \\
20 & apparent \textit{Swift} uvm2-band magnitude error \\
21 & apparent 2MASS $J$-band magnitude \\
22 & apparent 2MASS $J$-band magnitude error \\
23 & apparent 2MASS $H$-band magnitude \\
24 & apparent 2MASS $H$-band magnitude error \\
25 & apparent 2MASS $K$-band magnitude \\
26 & apparent 2MASS $K$-band magnitude error \\
27 & apparent UKIDSS $Y$-band magnitude \\
28 & apparent UKIDSS $Y$-band magnitude error \\
29 & apparent UKIDSS $H$-band magnitude \\
30 & apparent UKIDSS $H$-band magnitude error \\
31 & apparent UKIDSS $K$-band magnitude \\
32 & apparent UKIDSS $K$-band magnitude error \\
33 & $b/a$ axial ratio of outer disk \\
34 & $R_{50,r}$ half-light radius in $r$ band \\
35 & $R_{90,r}$ 90\% light radius in $r$ band \\
36 & $\Delta_{g-r}$ $g-r$ color gradient \\
37 & $(u-r)^m$ modeled $u-r$ color \\
38 & $(u-i)^m$ modeled $u-i$ color \\
39 & $(u-J)^m$ modeled $u-J$ color \\
40 & $(u-K)^m$ modeled $u-K$ color \\
41 & $(g-r)^m$ modeled $g-r$ color \\
42 & $(g-i)^m$ modeled $g-i$ color \\
43 & $(g-J)^m$ modeled $g-J$ color \\
44 & $(g-K)^m$ modeled $g-K$ color \\
45 & stellar mass \\
46 & foreground extinction in $u$ band \\
47 & foreground extinction in $g$ band \\
48 & foreground extinction in $r$ band \\
49 & foreground extinction in $i$ band \\
50 & foreground extinction in $z$ band \\
51 & foreground extinction in NUV band \\
52 & foreground extinction in uvm2 band \\
53 & foreground extinction in $Y$ band \\
54 & foreground extinction in $J$ band \\
55 & foreground extinction in $H$ band \\
56 & foreground extinction in $K$ band \\
\enddata
\tablecomments{All magnitudes are newly measured from the raw images. Apparent magnitudes are provided without foreground extinction corrections. Foreground extinction corrections used in this work are provided. Modeled colors designated by a superscript m are products of the SED fitting routine from K13, described in \S \ref{sec:mstar} and have foreground extinction corrections and k-corrections implicitly included. The datatable is provided at http://resolve.astro.unc.edu/data/resolve\_phot\_dr1.txt}
\label{tb:phottable}
\end{deluxetable}

\subsubsection{Comparison with Catalog Photometry}
\label{sec:compsdss}

\begin{figure*}
\epsscale{1.}
%\plotone{/srv/one/keckert/papers/govers/figs3/fig3.eps}
\plotone{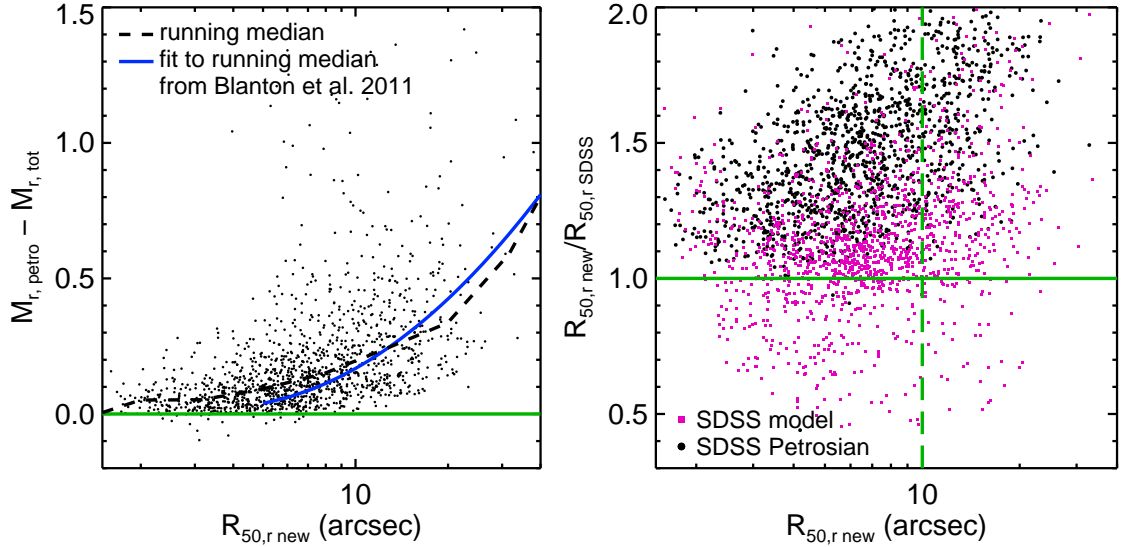}
\caption{Comparison of newly reprocessed photometry and SDSS DR7
  photometry.  a) Comparison of absolute Petrosian magnitudes
  M$_{r,petro}$ and absolute total magnitudes M$_{r,tot}$ from this
  work as a function of remeasured $R_{50,r}$. Our reprocessed
  photometry is brighter than the SDSS DR7 catalog Petrosian
  photometry; the running median is shown as a dashed black line. (We
  observe similar but slightly smaller offsets between our reprocessed
  photometry and the SDSS DR7 catalog model photometry.)  The solid
  blue line shows the expected magnitude difference using the new sky
  subtraction vs.\ the SDSS standard pipeline as a function of true
  galaxy $R_{50,r}$, based on the analysis of
  \citet{2011AJ....142...31B}, which is in excellent agreement with
  our data.  b) Comparison showing $R_{50,r}$ from our reprocessed
  photometry divided by $R_{50,r}$ from the DR7 photometry (Petrosian:
  black dots; model: pink squares), as a function of the new
  $R_{50,r}$ in arcsec.  The green solid line shows one-to-one
  correspondence, and the green dashed line marks 10\arcsec, above
  which the new background subtraction should significantly affect the
  measured flux and radius of the galaxy \citep{2011AJ....142...31B}.
  Our newly reprocessed photometry yields $\sim$13\% larger half-light
  radii than SDSS model half-light radii, with larger increases for
  galaxies with $R_{50,r}$ $>$ 10\arcsec. The increases over Petrosian
  radii are more extreme due to the assumption of circularity in the
  Petrosian algorithm.}

\label{fg:photcompmagrad}
\end{figure*}

We compare our newly reprocessed magnitudes, radii, and colors to the
Petrosian and model photometry provided in the SDSS DR7 catalog in
Figure \ref{fg:photcompmagrad}. SDSS catalog Petrosian magnitudes are
measured within a circular aperture of twice the Petrosian radius,
defined as the radius $R_{p}$ where the ratio of the local surface
brightness at $R_{p}$ to the surface brightness within $R_{p}$ is
equal to 0.2 \citep{2001AJ....121.2358B}. The SDSS pipeline uses the
Petrosian radius defined by the $r$ band to compute Petrosian
magnitudes for all other bands, thus yielding aperture-matched
magnitudes and colors. The Petrosian system should pick up nearly
total fluxes for disk (Sersic $n=1$) galaxies, but is known to
underestimate magnitudes for higher Sersic $n$ galaxies by $\sim$0.2
mag \citep{2005AJ....130.1535G}. The SDSS pipeline also computes model
magnitudes by fitting exponential ($n=1$) and de Vaucouleurs ($n=4$)
models to the galaxy light profile, choosing the model of greater
likelihood in the $r$ band, and extrapolating the profile to
infinity. To measure model magnitudes for the $ugiz$ bands, the SDSS
pipeline scales the amplitude of the $r$-band profile up or down to
best match the profile in that band
\citep{2002AJ....123..485S}. Neither magnitude system is ideal as the
Petrosian magnitudes do not measure the total galaxy light, while the
model magnitudes are most sensitive to the inner profile of the galaxy
and do not allow for color gradients within galaxies.

In Figure \ref{fg:photcompmagrad}a we compare our newly reprocessed
$r$-band magnitudes with DR7 Petrosian catalog magnitudes as a
function of galaxy half light radius $R_{50,r}$. The reprocessed
$r$-band magnitudes are overall brighter by \mbox{$\sim$0.13 mag} than
the DR7 Petrosian $r$-band magnitudes. We find a similar, but slightly
smaller, overall offset of \mbox{$\sim$0.1 mag} between our newly
reprocessed magnitudes and the DR7 model $r$-band magnitudes. The
offset increases for the largest galaxies, as seen in the running
median as a function of $R_{50,r}$ (black dashed line, Figure
\ref{fg:photcompmagrad}a). Much of this trend can be attributed to our
use of the improved sky background subtraction from
\citet{2011AJ....142...31B}, which was not available for DR7. The blue
solid line shows the expected median offset between galaxy magnitudes
using the new sky subtraction vs.\ the standard SDSS DR7 pipeline, as
a function of true galaxy $R_{50,r}$ (based on coefficients from Table
1 of \citealt{2011AJ....142...31B}, only valid for $R_{50,r}$ $>$
5\arcsec). Our running median matches very well with the expected
trend.  Note also that in this work, we do not use information from
the inner profile of the galaxy to compute extrapolated total
magnitudes, but rather extrapolate the light based on the outer
profile of the galaxy.  This difference may also contribute to the
generally brighter magnitudes that we measure.

In Figure \ref{fg:photcompmagrad}b we compare our newly remeasured
$R_{50,r}$ values with DR7 catalog $R_{50,r}$ values as a function of
the newly remeasured $R_{50,r}$. Since we measure greater flux per
galaxy, we expect the half-light radii to be larger, and indeed we
find that the new $R_{50,r}$ values are typically $\sim$49\% larger
than the SDSS Petrosian $R_{50,r}$ values and $\sim$13\% larger than
the model $R_{50,r}$ values. The ratio between the new and catalog
$R_{50,r}$ values becomes much greater above a remeasured $R_{50,r}$
of $\sim$10\arcsec, the value identified by
\citet{2011AJ....142...31B} as the true galaxy half light radius above
which the use of the new sky background should significantly affect
the measured galaxy flux and radius measurement. Another consideration
affecting only the Petrosian radii is the fact that SDSS Petrosian
apertures are circular whereas both our apertures and SDSS model
apertures are elliptical. When we restrict the comparison of half
light radii to galaxies with $b/a$ $>$ 0.85, our half light radii are
only $\sim$20\% larger than the Petrosian radii, more in line with the
13\% increase over the model radii. Figure 9 of
\citet{2012MNRAS.425.2741H} also shows the trend for Petrosian half
light radii to have greater disagreement with remeasured half-light
radii for more intrinsically edge-on galaxies.

\begin{figure*}
\epsscale{1.}
%\plotone{/srv/one/keckert/papers/govers/figs3/fig4.eps}
\plotone{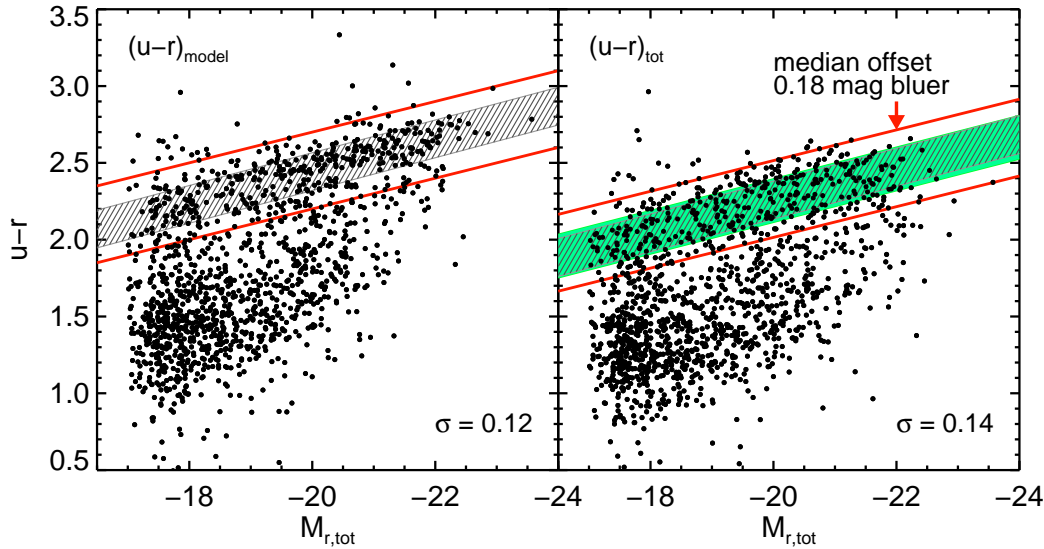}
\caption{Comparison of \cur{} color vs.\ our M$_{r,tot}$ for DR7 model
  colors (left) and our newly reprocessed colors (right). We determine
  the red sequence boundaries (red lines) for the DR7 data by eye,
  throwing out outliers that are too red.  We then fit a line to the
  data points between the two red lines and compute the rms which is
  0.12 for the DR7 model magnitudes (slope and width of red sequence
  shown by grey has marked region).  To do the same for the newly
  reprocessed photometry, we first determine the overall color offset
  by computing the median color of each distribution in bins of 0.2
  mag in M$_{r,tot}$.  We then find the median of the differences of
  the medians in each bin, obtaining an overall shift of 0.18 mag.  We
  shift down our red sequence boundaries by 0.18 mag for the newly
  reprocessed photometry.  We then fit a line and compute the rms to
  be 0.14 mag (slope and width shown in green solid region with slope
  and rms of DR7 model colors shifted down by 0.18 mags and
  overplotted), which is larger by 0.02 mag or 15\% than the rms
  computed for the DR7 colors. As argued in \S \ref{sec:compsdss}, we
  believe the higher scatter in our red sequence to be more correct,
  as our photometry does not suppress color gradients. Also most of
  the few extremely red outliers (including one at \cur$_{tot}$ = 3.6
  off the plot) are low surface brightness red galaxies whose shallow
  $u$ band data are the hardest to measure. A few others are those
  embedded galaxies for which we subtracted off the light of the
  larger galaxy.}

\label{fg:photcompcolor}
\end{figure*}

In Figure \ref{fg:photcompcolor} we compare the newly reprocessed
total \cur{} colors and the DR7 model \cur{} colors vs.\ M$_{r,
  tot}$. We find that the new \cur{} colors are overall $\sim$0.18 mag
bluer than the DR7 model \cur{} colors.  To compute this offset, we
measure the running medians of each color distribution as a function
of M$_{r,tot}$ in 0.2 mag bins and subtract the two sets of median
colors.  We then determine the median of these median color
differences, which is 0.18 mag.  A large portion of the offset is due
to the improved sky subtraction algorithm, but we also note the fact
that our newly reprocessed photometry allows for color gradients
whereas SDSS model colors do not. Galaxy color gradients have been
found in all galaxy types (e.g.,
\citealp{1996A&A...313..377D,2000ApJS..126..271J,2013ApJ...777..116C}). For
example in the Nearby Field Galaxy Survey, \citet{2000ApJS..126..271J}
find that early and late types have typical $B-R$ colors that are
bluer by 0.1-0.2 mag in their outer regions, while dwarf types
distribute evenly between blue and red color gradients. In this work
we have explicitly allowed for color gradients by computing the total
magnitudes in each band separately without the assumption of fixed
profile shape built into the SDSS model magnitude algorithm. Even the
Outer Disk Color Correction method fixes only the color outside the
A$_{out}$ annulus.

A consequence of ignoring color gradients is that the red sequence
defined by DR7 model \cur{} colors appears tighter than the red
sequence defined by our newly reprocessed \cur{} colors.  To quantify
the scatter, we fit a line to both sets of colors between the red
sequence boundaries marked off by the red lines in Figure
\ref{fg:photcompcolor}, and measure the rms from the fit. The red
sequence definition is shifted for the newly reprocessed \cur{} colors
by \mbox{0.18 mag} to account for their overall bluer colors.  We
confirm the visual impression that the DR7 red sequence is tighter,
finding that the SDSS model \cur{} red sequence is tighter by
$\sim$16\%.  This tight red sequence seems to be an artifact of the
SDSS model magnitude algorithm and should not be over-interpreted in
measuring the star formation histories of red sequence galaxies. We
note that \citet{2011ApJS..196...11S} also report that using separate
fits to compute $g$- and $r$-band magnitudes produces a more scattered
red sequence than obtained when fixing fits in both bands to have the
same half light radius, although these authors still choose fixed half
light radius fits for convenience.

\begin{figure*}
\epsscale{1.}
%\plotone{/srv/one/keckert/papers/govers/figs3/fig5.eps}
\plotone{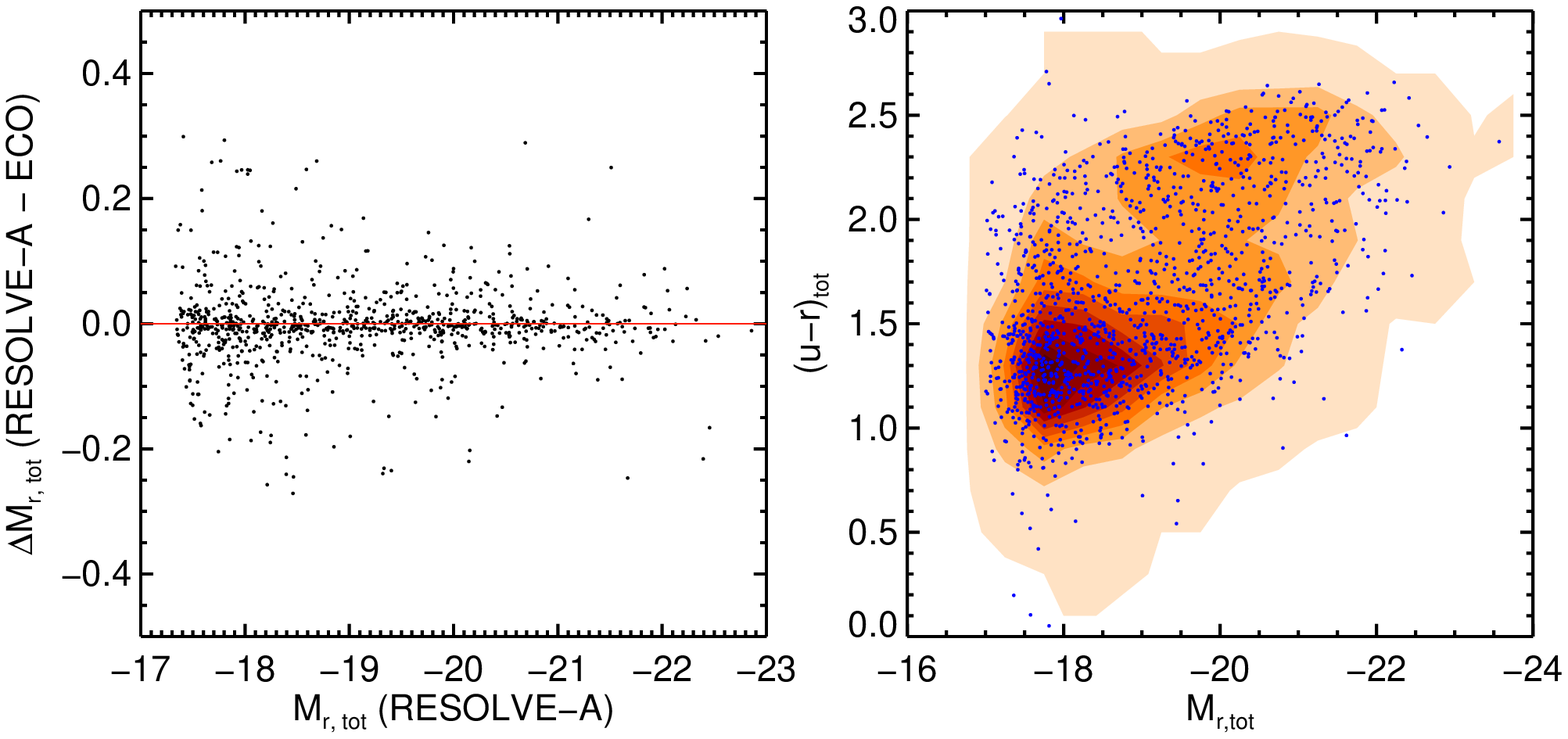}
\caption{Comparison of photometric measurements for galaxies in both
  the RESOLVE survey and the ECO catalog. a) Difference in M$_{r,tot}$
  measured for the same galaxies in RESOLVE-A and ECO vs.\ the
  RESOLVE-A M$_{r,tot}$. The main difference between the two
  measurements arises in the masking step. Every RESOLVE-A galaxy mask
  is checked by eye, but for the much larger ECO data set we check
  only images for which we have identified large discrepancies between
  extrapolated and aperture magnitudes. The two sets of measurements
  agree well, with differences mostly $<$0.2 mag.  Differences become
  larger for fainter galaxies, for which we expect larger
  uncertainties in extrapolation. b)~Color-magnitude relations of the
  full ECO (contours) and RESOLVE (dots) data sets, show that they are
  consistent. While we do not use ECO in this work, the ECO catalog an
  extension of the RESOLVE survey, so it is useful to establish that
  its photometry is consistent with the rest of RESOLVE.}

\label{fg:compareweco}
\end{figure*}

In Figure \ref{fg:compareweco} we compare independent photometric
measurements for RESOLVE survey galaxies that overlap with the ECO
(Environmental COntext) catalog (Moffett et al., submitted), which is
a larger volume-limited data set encompassing the RESOLVE-A subvolume.
The ECO catalog has been reprocessed through the same pipeline, with
the most significant difference in methodology occurring at the
masking step.  Since ECO has \mbox{$\sim$10} times the number of
galaxies as RESOLVE, for ECO it was not feasible to check each mask by
hand.  The most egregious cases of over- or under-masking were
determined in a preliminary run of the photometry code on the catalog,
by checking for extrapolated magnitudes that significantly disagreed
with aperture magnitudes or cases where no magnitude was measured.
The masks for these galaxies were then checked by eye to mitigate
under-/over-masking.  We compare the ECO and RESOLVE M$_{r,tot}$
measurements for galaxies in the overlapping subvolume in Figure
\ref{fg:compareweco}a.  We find no offset and only small differences
of typically $<$0.2 mag between the two sets of magnitudes. Some of
these differences may be attributed to the final magnitude chosen by
the pipeline (Outer Disk Fit or Curve of Growth for the $r$ band),
which is based on the degree to which the frame is masked.  Figure
\ref{fg:compareweco}b shows the color-magnitude plots for both the
full RESOLVE-A and RESOLVE-B (blue points) and ECO (orange-red
contours) data sets, demonstrating that they are consistent.

An independent validation of the methods used in this work is shown in
Figure 2a of K13 for the Nearby Field Galaxy Survey
\citep{2000ApJS..126..271J}. All NFGS galaxies with available SDSS
data were reprocessed through the same pipeline as described here and
Figure 2a in K13 shows that the reprocessed \cur{} colors are
consistent with the expected Vega-AB offset for $U-R$ colors measured
in \citet{2000ApJS..126..271J} over all angular sizes.  The comparison
between the total \cur{} color and the SDSS DR7 model \cur{} colors
reveals an offset such that the new photometry yields $\sim$0.2 mag
bluer \cur{} colors than the DR7 model colors, similar to the offset
that we measure.

%We note that the three methods used, even the color correction method
%which fixes the color in the outer disk, more correctly model color
%gradients in galaxies than the model magnitudes from SDSS.  The model
%magnitudes are based on the model of higher likelihood (exponential or
%deVaucouleurs) determined for the $r-$band.  The $r-$band model is
%then scaled in amplitude to fit to the $ugiz$ bands
%\citep{2002AJ....123..485S}.

\subsection{Stellar Masses}
\label{sec:mstar}

Stellar masses and k-corrected colors are calculated using the
spectral energy distribution (SED) modeling code described in
\citet{2007ApJ...657L...5K}, as modified by K13, which fits a grid of
stellar population models to our newly reprocessed total
NUV$ugrizYJHK$ magnitudes plus new \textit{Swift} uvm2 data for 19
galaxies). With photometric data from up to 10 bands, we are able to
estimate robust stellar masses. We omit UKIDSS $YHK$ values if the
frames have been flagged by eye. We also omit UKIDSS $HK$ and 2MASS
$JHK$ if the values are fainter than 18, 17.5 and 16, 15, 14.5
respectively. We also remove any NUV magnitudes fainter than 24, and
we remove the $u$ band magnitudes for four galaxies for which the $u$
band data available from SDSS are essentially frames of noise.

In this work use the second model grid from K13, which is a grid of
composite stellar population models (CSPs) including an old simple
stellar population (SSP) ranging in age from 2-12 Gyr and a young
population either described by continuous star formation starting 1015
Myr ago and turning off between 0 to 195 Myr ago or as a quenching
burst with SSP age 360, 509, 641, 806, or 1015 Myr. The contribution
from the young population ranges from 1-94.1\% of the stellar mass.
The model grid is built using the stellar population models from
\citet{2003MNRAS.344.1000B} with a Chabrier IMF
\citep{2003PASP..115..763C}, and four possible metallicities (Z =
0.004, 0.008, 0.02, or 0.05). Eleven reddening values
  ($\tau$v ranges from \mbox{0--1.2}) are applied to the young
  population following the dust extinction law given by
  \citet{2001PASP..113.1449C}. There is no physical or spatial model
  assumed for the dust, only an empirical determination of the amount
  of reddening and extinction based on the stellar population model
  grid fits to the galaxy SED.

To determine a galaxy's stellar mass, the stellar mass is computed for
each CSP model in the grid and given a likelihood based on the $\chi^2$ value
of the model fit to the data. Combining likelihoods over all models
yields a stellar mass likelihood distribution for each galaxy.  The
median value of this stellar mass likelihood distribution is taken to
be the nominal stellar mass of the galaxy.

\begin{figure}
\epsscale{1.}
%\plotone{/srv/one/keckert/papers/govers/figs3/fig6.eps}
\plotone{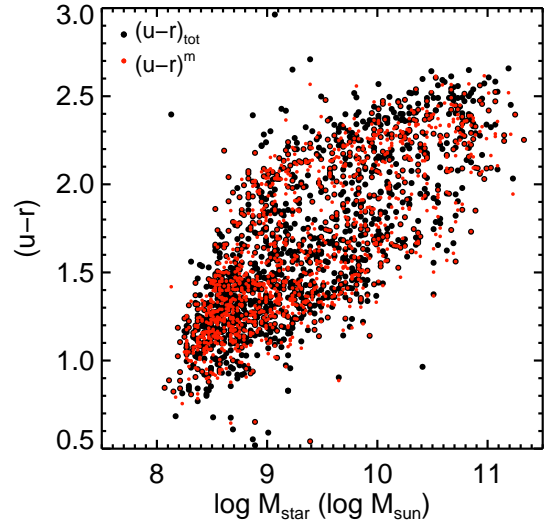}
\caption{Color vs.\ stellar mass for the entire RESOLVE survey.  Black
  dots show total reprocessed magnitudes \cur{} and smaller red dots
  show the SED modeled colors \curm.  While there are outliers in the
  reprocessed photometry as seen in Figure \ref{fg:photcompcolor}, the
  SED modeled colors use information over the entire NUV$ugrizYJHK$
  SED and provide a cleaner color-mass plot.}

\label{fg:colormass}
\end{figure}

The SED modeling code also outputs the likelihood weighted colors for
each galaxy, which are effectively ``smoothed'' by the model fits and
implicitly k-corrected.  We denote the use of these modeled colors
with a superscript $m$ (following the notation from K13 and Moffett et
al., submitted). The stellar population code also outputs de-extincted
galaxy magnitudes, taking into account the internal extinction due to
dust in the galaxy. These magnitudes cleanly divide the red and blue
sequences in the color-stellar mass diagram as shown in K13 and
Moffett at al., submitted.  Here, however, we choose to use the
modeled colors, which represent the actual rest-frame colors of the
galaxies, for easier application of the PGF calibrations to other data
sets.  We show the color-stellar mass diagram for RESOLVE-A and
RESOLVE-B using both \cur{} total colors measured from the raw
reprocessed photometry and \curm{} colors from the model fits in
Figure \ref{fg:colormass}.  The SED modeled colors agree with the
measured colors well within the expected k-correction values at these
redshifts of up to 0.03 mag for the $r$ band and 0.1 mag for the $u$
band.

Since some stellar mass estimation techniques have been shown to be
biased as a function of inclination \citep{2009ApJ...691..394M}, it is
important to test whether our stellar masses may be biased as a
function of axial ratio. We perform two tests. First, we select only
galaxies with gas-to-stellar mass ratio $>$ 0.1, implying significant
gas and thus potentially dust, and we divide this subset into
quartiles based on their axial ratio. A KS test reveals that the
stellar masses of the upper and lower quartiles ($b/a$ $>$ 0.77 and
$b/a$ $<$ 0.39) are consistent with being drawn from the same
population ($p_{null} = 0.99$). Second, we recompute our stellar
masses applying the dust law to \textit{both} the young and old
populations (as opposed to just the young population as for our
preferred mass estimates). We find a tiny overall offset for late type
galaxies of $\sim$0.02 dex but no systematic trend between the two
stellar mass calculations as a function of axial ratio. For early type
galaxies we find a tiny differential systematic offset of 0.02 dex
between the most elongated and roundest galaxies. Both of these tests
suggest our stellar mass calculations are not biased by dust
extinction.

\subsection{HI Masses}
\label{sec:himasses}

The HI masses and upper limits for RESOLVE come from the blind 21cm
ALFALFA survey \citep{2011AJ....142..170H} and our own new
observations with the GBT and Arecibo telescopes. It is important for
creating a gas mass estimator to have complete HI data for the entire
data set. Below we describe the observations taken to date, how we
determine and handle confused sources, and the HI completeness of the
RESOLVE data set.

The ALFALFA survey has covered the entire RESOLVE-A region and the Dec
0$^{\circ}$ to +1.25$^{\circ}$ strip of RESOLVE-B, providing HI
detections or upper limits (not necessarily strong) for 85\% of
RESOLVE. Data reduction and source extraction are described in
\citet{2011AJ....142..170H}. At the nominal S/N limit of 6, the
ALFALFA flux limit translates to a fixed HI mass sensitivity at
RESOLVE distances of $\sim$10$^{9}$ \msun.  Since RESOLVE galaxies
range from 10$^{9}$--10$^{11.5}$ \msun, this fixed sensitivity implies
a large number of upper limits that are much weaker than our stated
goal of 1.4M$_{HI}$ $<$ 0.05M$_{star}$.  To increase the yield from
the basic ALFALFA data products, Stark et al.\ (in prep.) extract
140 lower S/N detections and upper limits for RESOLVE
galaxies within the ALFALFA grids.

To further increase the useful HI data set, we have acquired pointed
observations with the GBT and Arecibo telescopes obtaining HI data for
290 galaxies in RESOLVE-A and 337 galaxies in RESOLVE-B (Stark et
al.\ in prep.).  We target galaxies with either no HI measurements or
weak upper limits from ALFALFA, aiming for detections with S/N $\sim$
10 or strong upper limits. In addition, we have cross-matched the
RESOLVE catalog with the HI catalog of \citet{2005ApJS..160..149S} to
obtain thirteen more HI measurements.

To check for consistency between our GBT and Arecibo pointed
observations, we have remeasured HI fluxes for $\sim$10 galaxies in
RESOLVE and find consistency between observations with the two
telescopes within $\sim$15--20\% (Stark et al.\ in prep.).  Consistency
checks between ALFALFA and Arecibo pointed observations from the
\citet{2005ApJS..160..149S} catalog are documented in
\citet{2011AJ....142..170H} and HI flux measurements between the two
catalogs are shown to be in agreement within $\sim$20\%.

HI masses and upper limits are calculated as described in Stark et
al.\ (in prep.) and K13. Confusion is determined based on the source of
the HI measurement, 4\arcmin{} for the smoothed resolution element of
ALFALFA, 9\arcmin{} for the GBT, and 3.5\arcmin{} for Arecibo pointed
observations. De-confusion is performed following techniques described
in Stark et al.\ (in prep) that improve on methods described in K13. The
de-confused HI masses are provided in Stark et al.\ (in prep.). Neutral
gas masses are calculated by multiplying the HI mass by 1.4 to account
for helium, M$_{gas}$ = 1.4M$_{HI}$.

For this work, we apply the following set of criteria to determine
reliable HI masses. We require detections to have S/N $>$ 5. We use
de-confused HI masses if the systematic error on the de-confused HI
masses is $<$ 25\% of the de-confused HI mass. For limits, we require
that the upper limit yields a gas mass $<$ 0.05M$_{star}$.

Based on these criteria, we provide the statistics on HI completeness
for this work for the two RESOLVE subvolumes. RESOLVE-A has a total of
955 galaxies, of which 637 have reliable HI detections with S/N $>$ 5
(34 of those detections are successfully deconfused observations) and
107 have strong upper limits resulting in M$_{gas}$ $<$
0.05M$_{star}$. Thus 78\% of the sample (744 galaxies) have reliable
HI data for defining PGF calibrations.  In RESOLVE-B there are 487
galaxies, of which 294 have good detections (34 are successfully
deconfused observations) and of which 70 are strong upper limits,
yielding 75\% of RESOLVE-B or 364 galaxies that have reliable HI data.

% Of
%the remaining 124 galaxies, 3 are low S/N detections, 72 are weak
%upper limits, 38 are cases where it is impossible to de-confused the
%HI data, and 12 have no HI observation to date.

%, limiting the sample at M$_{r, tot}$ $<$ -17.33, 68\% of
%the galaxies have good HI detections that are either not confused or
%successfully de-confused.  Another 12\% are strong upper limits
%resulting in M$_{gas}$ $<$ 5\%$\times$M$_{star}$. The RESOLVE-A data
%set is then 80\% complete with good HI masses and upper limits.  Of
%the remaining 20\% of galaxies, 12\% are due to weak upper limits and
%8\% are objects where confusion is impossible to disentangle.

%In RESOLVE-B, limiting the sample at M$_{r, tot}$ $<$ -17.0, 62\% of
%the galaxies have good HI detections that are either not confused or
%successfully de-confused.  Another 13\% are strong upper limits
%resulting in M$_{gas}$ $<$ 5\%$\times$M$_{star}$. The RESOLVE-B data
%set is then 75\% complete with good HI masses and upper limits.  Of
%the remaining 25\% of galaxies, 16\% are due to weak upper limits and
%6\% are objects where confusion is impossible to disentangle and 3\%
%have no observations.

\begin{figure}
\epsscale{1.}
%\plotone{/srv/one/keckert/papers/govers/figs3/fig7.eps}
\plotone{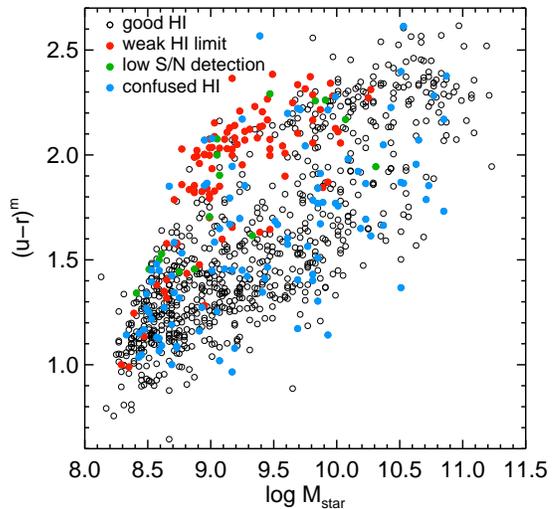}
\caption{Color vs.\ stellar mass relationship for the RESOLVE-A data
  set.  Black open circles show galaxies with reliable HI detections
  as described in \S \ref{sec:himasses}.  Red dots show galaxies with
  HI upper limits yielding M$_{gas}$ $>$ 0.05M$_{star}$, green dots
  show galaxies with low S/N detections (S/N $<$ 5), and blue dots
  show galaxies with HI profiles that are impossible to
  de-confuse. The weak upper limits tend to be the low-mass red
  galaxies, which are generally gas poor and have the least absolute
  HI content of galaxies in RESOLVE. The low S/N galaxies are also low
  mass, but generally bluer.  The confused galaxies are interspersed
  throughout color and stellar mass.}

\label{fg:colormassforHI}
\end{figure}

In RESOLVE-A, we are still lacking adequate HI measurements for 211
galaxies.  Of those 211 galaxies, 92 are weak upper limits yielding
gas masses that range from 0.052 - 4.01 $\times$ M$_{star}$, 16 are
low S/N detections, and 103 have HI profiles where de-confusion is not
possible. To examine whether these galaxies with inadequate HI
measurements are biased, we plot color vs.\ log(M$_{star}$) for
RESOLVE-A in Figure \ref{fg:colormassforHI}. Galaxies for which we
have weak upper limits (red dots) tend to be low-mass, red galaxies
which are generally gas-poor and require the longest observing times
for successful detection or strong enough upper limits. Galaxies for
which we have low S/N detections (green dots) are low-mass blue
objects or higher mass red objects. Galaxies for which de-confusion is
impossible (blue dots) are scattered throughout color and stellar
mass.

\section{Color-limited PGF Calibrations}
\label{sec:simplecals}

\begin{figure}
%\plotone{/srv/one/keckert/papers/govers/figs3/calfigs/cal.u-Jfit.new.eps}
\plotone{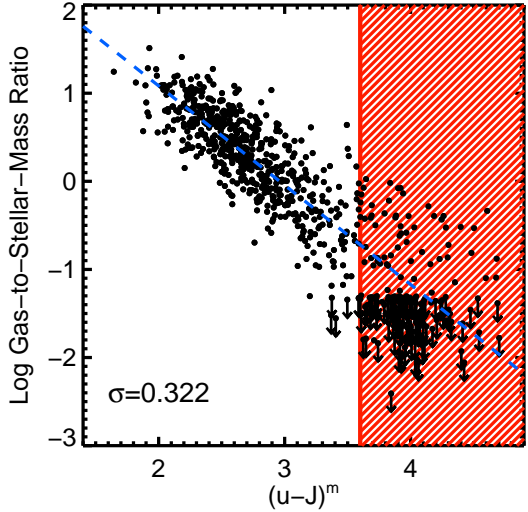}
% made with /srv/one/keckert/NFGS_DR8/figures/cal.plot3rd.pro
\epsscale{1.0}
\caption{Photometric gas fractions relation for \cuj{} SED modeled
  color, hereafter \cujm. For the photometric gas fractions relation
  gas refers to the atomic gas content, for which we include a helium
  correction factor: M$_{gas} = 1.4$M$_{HI}$. When fitting, we exclude
  galaxies redder than \cujm{} = 3.6 mag, where the
  quenched galaxy population overwhelms the low log(G/S) end of the
  relationship between log(G/S) and color. We also exclude galaxies
  bluer than 2.0 mag as there are only a few data points that may skew
  the overall fit.}

\label{fg:gslincal}
\end{figure}

In this section we describe our method to provide z=0 PGF calibrations
via linear fits between log(G/S) and color.  In Figure
\ref{fg:gslincal}, we show the relationship between log(G/S) and
\cujm{} color, which is clearly linear. However, for galaxies redder
than \cujm{} = 3.6 mag there is a breakdown in the correlation. While
the correlation between log(G/S) and \cujm{} color continues for some
galaxies redder than 3.6 mag, we also see that the population of
quenched galaxies with very low values of log(G/S) becomes more
important for these same red colors. In \S \ref{sec:cals}, we describe
a new calibration method using a 2D model fit to the probability
density field of log(G/S) vs.\ color that allows us to model all
galaxies.

Here we are generating linear fits to predict values of log(G/S) from
color, where the latter has much smaller errors and thus functions as
a classical independent variable. Thus, especially given the
likelihood of the intrinsic scatter over and above the errors, to
obtain the best predictor we should minimize residuals in log(G/S)
alone \citep{1990ApJ...364..104I,1992ApJ...397...55F}. Due to the
population of red galaxies with strong HI upper limits, we must
exclude all galaxies redward of a vertical color cutoff, e.g., \cujm{}
$>$ 3.6 mag, similar to \citet{2012A&A...544A..65C} and K13. Making a
cut in color is appropriate for measuring the correct calibration to
predict log(G/S) from color as we want to preserve the scatter for the
predicted quantity \citep{2002AJ....123.2358K} rather than fitting to
only the HI detections or making a cut in log(G/S). Excluding red
galaxies, however, limits the validity of our PGF calibration to
galaxies blueward of the red color cutoff.

%One tactic we can use to obtain a PGF calibration is to exclude all
%galaxies redward of a vertical color cutoff, e.g., \cujm{} $>$ 3.6
%mag, similar to \citet{2012A&A...544A..65C} and K13.  Making a cut in
%color is appropriate for measuring the correct calibration to predict
%log(G/S) from color as we want to preserve the scatter for the
%predicted quantity \citep{2002AJ....123.2358K}. When fitting a line to
%predict a quantity, we want to minimize that quantity's 
%scatter. To create the best PGF calibration then, we must use a color
%cut rather than fitting to only the HI detections or making a cut in
%log(G/S). Excluding red galaxies, however, limits the validity of our
%PGF calibration to galaxies blueward of the red color cutoff.

A set of such color-limited PGF calibrations for a variety of color
combinations is summarized in Table \ref{tb:simplecals}. To create
these, we use the 744 galaxies from the RESOLVE-A
data set that have reliable HI detections or strong upper limits.
``Reliable'' HI detections are considered to include non-confused
detections with S/N $>$ 5 as well as de-confused detections where the
systematic error on the deconfused gas mass is $<$ 25\% of the
measured gas mass (see \S \ref{sec:himasses}).  We define strong upper
limits to be those for which the gas mass is $<$ 5\% of M$_{star}$.
We exclude galaxies redder than the red color cutoff of each color
distribution listed in Table \ref{tb:simplecals} (roughly where the
upper limits start to dominate). We also trim points at the blue end
where the density of points is low and outliers may affect the fit as
indicated in Table \ref{tb:simplecals}.  For \cujm{} color, the blue
color trim is 2.0 mag and the red color cutoff is
3.6 mag.  Finally we perform an ordinary least
squares forward fit to minimize the scatter in log(G/S), the quantity
that we want to predict. We choose not to weight the
  fit by the measurement uncertainties in log(G/S), because they are
  correlated with the values of log(G/S) and color, so weighting by
  them would bias the fits towards galaxies with high gas content.
The slope and offset in log(G/S) of these color-limited PGF
calibrations are given in Table \ref{tb:simplecals} along with the
measured scatter in the relations and the blue color trim and red
color cutoff values.\footnote{We note that the RESOLVE-A region, which
  we use for these linear fits, is less redshift complete than
  RESOLVE-B. We have performed empirical completeness corrections
  based on luminosity and surface brightness or color for the ECO
  catalog (Moffett et al., submitted), which encompasses the RESOLVE-A
  subvolume. These empirical completeness corrections are based on the
  more complete RESOLVE-B subvolume. We find that weighting the linear
  fits by these completeness corrections does not change the linear
  fit parameters significantly and we do not use the completeness
  corrections in this work.}

\begin{deluxetable*}{ccccccc}
\tablecaption{Color-limited Photometric Gas Fraction Calibrations}
\tablehead{\colhead{color} & \colhead{slope} & \colhead{log(G/S) offset} & \colhead{$\sigma$} & \colhead{blue trim} & \colhead{red cutoff} & \colhead{N galaxies}\\ \colhead{(mag)} & \colhead{(dex/mag)} & \colhead{(dex)} & \colhead{(dex)} & \colhead{(mag)} & \colhead{(mag)} &\colhead{}} 
\startdata
$(u-r)^m$ & -1.763 & 2.725 & 0.319 & 1.0 & 2.0 & 552 \\ 
$(u-i)^m$ & -1.421 & 2.510 & 0.314 & 1.0 & 2.3 & 571 \\ 
$(u-J)^m$ & -1.127 & 3.337 & 0.322 & 2.0 & 3.6 & 560 \\ 
$(u-K)^m$ & -1.059 & 3.993 & 0.331 & 2.8 & 4.4 & 543 \\ 
$(g-r)^m$ & -3.488 & 1.467 & 0.302 & 0.1 & 0.6 & 568 \\ 
$(g-i)^m$ & -2.399 & 1.546 & 0.310 & 0.2 & 0.9 & 557 \\ 
$(g-J)^m$ & -1.582 & 2.918 & 0.332 & 1.1 & 2.2 & 550 \\ 
$(g-K)^m$ & -1.401 & 3.744 & 0.364 & 2.0 & 3.0 & 501 \\ 
\enddata
\label{tb:simplecals}
\end{deluxetable*}

We expect these fits to be useful for galaxies blueward of the blue
color trim (as discussed in \S \ref{sec:discussion}, but these
calibrations do not allow us to predict gas masses for galaxies redder
than the red color cutoff. We also note that color-limited linear
calibrations, and all calibrations based on simple fits, are subject
to bias without survival analysis to model galaxies that are confused,
have weak upper limits, or lack reliable HI detections. Routines to
incorporate upper limits in a linear fit exist but rely on the
assumption that the upper limits are distributed randomly throughout
the sample (as discussed in \citealt{1986ApJ...306..490I}). Such
``random censoring'' is not the case for the PGF calibration, as those
galaxies with upper limits in the RESOLVE-A data set are primarily red
galaxies with low gas-to-stellar mass content. Thus using such
routines would not be statistically robust. In \S \ref{sec:cals} we
improve on simple fitting by producing a 2D model of the log(G/S)
vs.\ color probability density field. This fully probabilistic
approach allows us to implement a version of survival analysis that
reinserts the galaxies left out of the linear fits and to predict
log(G/S) probability distributions for individual galaxies, even those
redder than the color cutoff.  Before performing this analysis,
however, we analyze whether the residuals from these linear fits
correlate with any other photometric parameters that may help produce
tighter PGF relations.

\section{Correlations with 3rd Parameters}
\label{sec:3pcorr}

In this section, we seek a combination of color and other photometric
parameters that may produce a tighter PGF calibration for more
accurate gas mass estimation. To this end we use the RESOLVE-A data
set to analyze correlations between various photometric parameters and
residuals from the color-limited PGF calibrations in \S
\ref{sec:residexam}. We then explore possible physical reasons for
these residual correlations by examining their relation to galaxy
morphology in \S \ref{sec:residphys}. Lastly we provide plane fits
between color, axial ratio, and log(G/S) for tighter color-limited PGF
calibrations in \S \ref{sec:modcolpgfcal}.

\subsection{Best 3rd Parameter for Gas Mass Estimation}
\label{sec:residexam}

\begin{figure*}
\epsscale{0.9}
%\plotone{/srv/one/keckert/papers/govers/figs3/calfigs/residualsnew.u-Jfit.new.eps}
\plotone{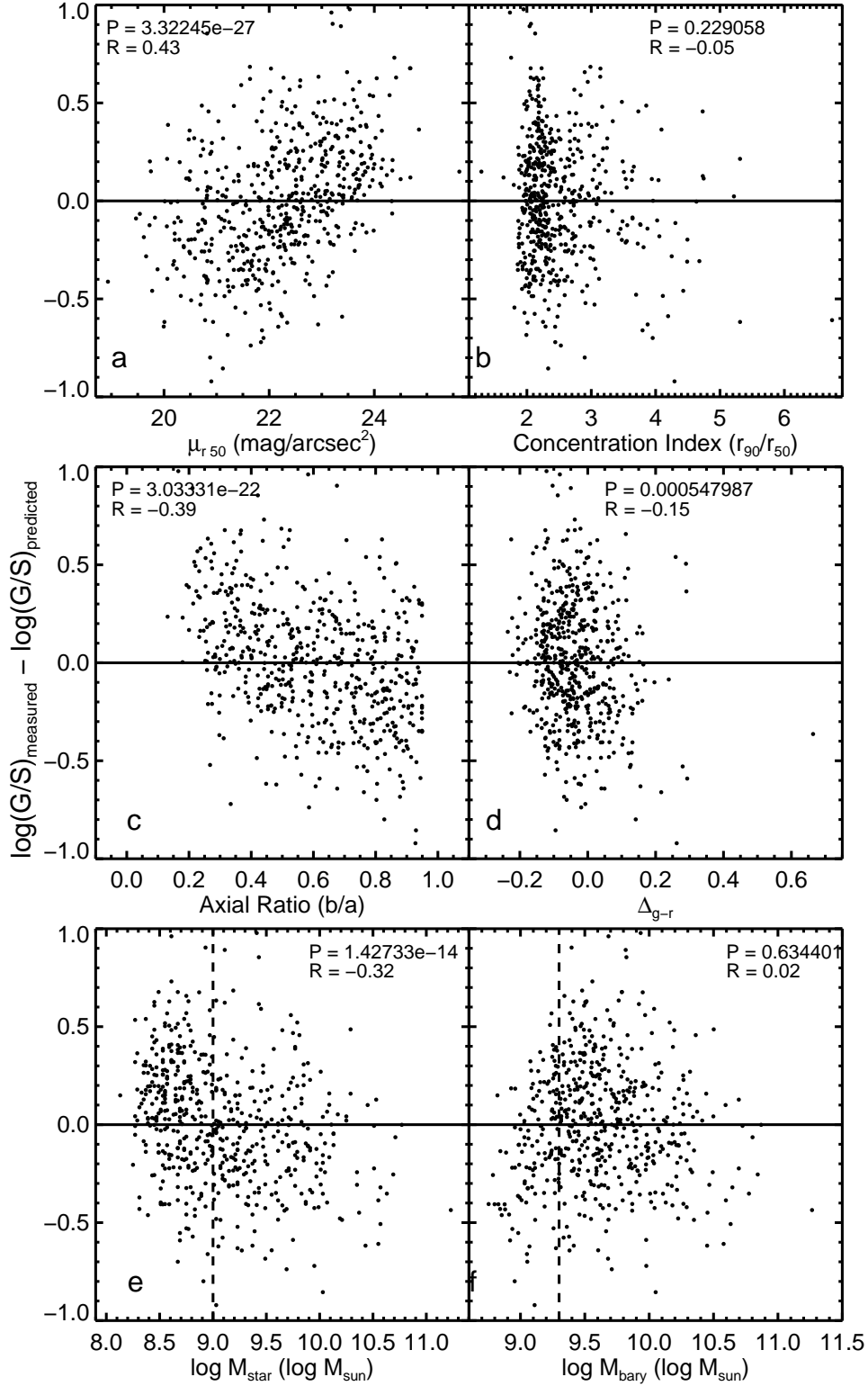}
\caption{Measured minus predicted log(G/S) for RESOLVE-A from the PGF
  calibration for \cujm. Residuals are plotted against a) $r$-band
  50\% surface brightness $\mu_{r,50}$, b) concentration index $C_r$
  ($R_{90}/R_{e}$), c) axial ratio $b/a$, d) color gradient
  $\Delta_{g-r}$, e) stellar mass M$_{star}$, and f) baryonic mass
  M$_{bary}$. We use the Spearman Rank test to assess whether a
  correlation exists between the residuals in log(G/S) and these
  photometric parameters.  The probability of no correlation and the
  strength of each correlation are reported for each panel. The
  parameters $\mu_{r 50}$, $\Delta_{g-r}$, and $b/a$ have significant
  correlations with the residuals in log(G/S). The correlation with
  M$_{star}$ is not physically meaningful but is caused by covariance
  as well as our selection on M$_{r,tot}$, a proxy for baryonic mass,
  which implies that the data set contains low stellar mass gas-rich
  objects but not low stellar mass gas-poor objects. We mark in panel
  e the stellar mass completeness limit for RESOLVE-A (M$_{bary}$ =
  9.0), below which the residuals are clearly biased towards higher
  log(G/S) measured than predicted. We also mark in panel f the
  baryonic mass completeness limit for RESOLVE-A (M$_{bary}$ =
  9.3). Below this limit we lack extremely gas-dominated residuals,
  which is expected due to the absolute magnitude limited sample,
  where the highest high baryonic mass-to-light ratio will fall below
  our M$_{r,tot}$ = $-17.33$ mag limit.}

\label{fg:gsresidall}
\end{figure*}

As potential third parameters, we examine the surface brightness
within $R_{50,r}$ ($\mu_{r,50}$), concentration index ($C_r$),
photometric axial ratio ($b/a$), \cgr{} color gradient
($\Delta_{g-r}$), stellar mass M$_{star}$, and baryonic mass
M$_{bary}$. The parameter $\mu_{r,50}$ is found by determining
$R_{50,r}$ from the ellipse profiles produced in the photometric
reprocessing. The $r$-band light within $R_{50,r}$ is then divided
over a circular area defined by $R_{50,r}$, giving a somewhat
intrinsic surface brightness for each object, although we have not
applied a correction for internal galaxy extinction or optical depth
since the goal is to provide a simple empirical recipe for gas
estimation. Concentration index is defined as $R_{90,r}$/$R_{50,r}$
following \citet{2001AJ....122.1238S} and
\citet{2001AJ....122.1861S}. The $b/a$ measurement comes from the
fixed ellipse fits in our photometric pipeline. $\Delta_{g-r}$ is
defined as the \cgr{} color within the annulus between the $r$-band
50\% and 75\% light radii minus the \cgr{} color within the $r$-band
half light radius, with more positive numbers meaning the galaxy has a
bluer center following \citet{2004AJ....127.1371K}. The stellar mass
M$_{star}$ is the median stellar mass from the likelihood weighted SED
models described in \S \ref{sec:mstar} and the baryonic mass is
M$_{star}$ + M$_{gas}$.

Figure \ref{fg:gsresidall} shows log(G/S) residuals from the \cujm{}
color-limited PGF calibration plotted against these different
parameters.  Only the data from galaxies used in the linear fit are
shown, i.e., those having a color between the blue trim and red cutoff
colors and reliable HI data or a strong HI upper limit. We have
performed Spearman Rank tests to assess whether there is a correlation
between the residuals in log(G/S) and these third parameters and
quantify the strength of that correlation. Third parameters
$\mu_{r,50}$, $b/a$, $\Delta_{g-r}$, and M$_{star}$ all show some
correlation with residuals in log(G/S), while $C_r$ and M$_{bary}$
show no significant correlation. Below we examine each significant
correlation to determine the best parameter for more accurate gas mass
prediction.

The correlation between residuals in log(G/S) and M$_{star}$, shown in
Figure \ref{fg:gsresidall}e, is in the sense that low stellar mass
objects tend to have higher measured log(G/S) than predicted, which is
expected from covariant errors in log(G/S) and
M$_{star}$. Furthermore, the bias becomes very evident for stellar
mass galaxies $<$10$^{9.0}$ \msun{} (roughly the limiting stellar mass
completeness limit) marked in panel e by a dashed line. This
correlation between log(G/S) residuals and stellar mass is caused by
our survey selection on M$_{r,tot}$, which serves as a close proxy for
baryonic mass (\citealp{2008AIPC.1035..163K}, K13) rather than stellar
mass. For a volume-limited, baryonic mass limited data set, an
abundance of gas-rich low stellar mass objects is built in.  The
correlation between log(G/S) residuals and stellar mass is thus not
surprising and only reflects the $r$-band magnitude selection.  In
Figure \ref{fg:gsresidall}f we see that the color-limited PGF
calibration does not show evidence for a correlation between residuals
in log(G/S) and baryonic mass.

The remaining correlations between log(G/S) residuals and photometric
parameters that we investigate are $\mu_{r,50}$, $b/a$, and
$\Delta_{g-r}$. These correlations may be related to galaxy morphology
and/or evolutionary state, which we explore in Figure
\ref{fg:2dresids} and \S \ref{sec:residphys}. To decide which of these
third parameters is best for use in our PGF calibrations, we use the
following criteria: 1) the third parameter has a strong correlation
with residuals in log(G/S), 2) the correlation between the third
parameter and log(G/S) residuals is not covariant with the built in
correlation between residuals in log(G/S) and stellar mass, and 3) the
third parameter is a reliable, well determined quantity for all
galaxies in the data set.

First we examine $\mu_{r,50}$ in Figure \ref{fg:gsresidall}a. The
correlation between residuals in log(G/S) and $\mu_{r,50}$ is such
that lower surface brightness galaxies have larger log(G/S) values
than predicted. The Spearman Rank test shows that the strength of the
correlation between residuals in log(G/S) and $\mu_{r,50}$ is high (R
= 0.43) with small probability of no correlation (P =
3$\times$10$^{-27}$). To test whether stellar mass is impacting the
correlation between residuals in log(G/S) and $\mu_{r,50}$, we fit a
line to the correlation between residuals in log(G/S) as a function of
stellar mass, and remove the stellar mass dependence from the
residuals. We then run a Spearman Rank test on the correlation between
the stellar mass corrected residuals in log(G/S) and $\mu_{r,50}$,
finding that the strength decreases to R = 0.28 and the probability of
no correlation is larger by several orders of magnitude (P =
9$\times$10$^{-12}$) though still significant. The reliability of
$\mu_{r,50}$ values is variable since $\mu_{r,50}$ is evaluated within
the half light radius. For the smallest galaxies, the quantity may not
truly represent the surface brightness within $R_{50,r}$.

Next we examine the third parameter $b/a$ shown in Figure
\ref{fg:gsresidall}c. The correlation between residuals in log(G/S)
and $b/a$ is such that more edge-on or disky galaxies have larger
log(G/S) values than predicted. The strength of the correlation
between residuals in log(G/S) and $b/a$ is also high (R = $-0.39$)
with small probability of no correlation (P =
3$\times$10$^{-22}$). Running a Spearman Rank test on the correlation
between the stellar mass corrected residuals in log(G/S) and $b/a$, we
find that both the correlation strength and the probability of no
correlation remain mostly the same (R = $-0.37$ and P =
5$\times$10$^{-20}$) showing that the correlation is not affected by
M$_{star}$. The $b/a$ measurements are reliable, since they are
evaluated over the outer disk of the galaxy. For the smallest galaxies
there may be a tendency to measure rounder objects, but since $b/a$ is
evaluated in the outer disk, this issue should be minimized.

Lastly we consider the third parameter $\Delta_{g-r}$ shown in Figure
\ref{fg:gsresidall}d. The correlation between residuals in log(G/S)
and $\Delta_{g-r}$ is such that galaxies with larger values of
$\Delta_{g-r}$ (more blue centered) have lower log(G/S) than
predicted. The strength of the correlation between residuals in
log(G/S) and $\Delta_{g-r}$ is much lower than the other two
parameters (R = $-0.15$) with a larger probability of no correlation
(P = 5$\times$10$^{-4}$). We test whether stellar mass is impacting
this correlation between residuals in log(G/S) and $\Delta_{g-r}$, by
removing the correlation between residuals in log(G/S) and stellar
mass. Running a Spearman Rank test on the correlation between the
stellar mass corrected residuals in log(G/S) and $\Delta_{g-r}$, we
find that both the correlation strength becomes stronger (R = $-0.25$)
and the probability of no correlation becomes several orders of
magnitude smaller (P = 8$\times$10$^{-10}$). The correlation between
$\Delta_{g-r}$ and stellar mass has been shown previously in
\citet{2013ApJ...769...82S} and it is apparent that the stellar mass
affects how this third parameter relates to residuals in log(G/S). The
reliability of the measurement may be suspect for the smallest
galaxies, since it requires measuring colors within $R_{50,r}$ and
between $R_{50,r}$ and $R_{75,r}$.

The quantity that best meets our three criteria is $b/a$ since it
exhibits a strong correlation with the residuals in log(G/S) that is
not affected by the correlation with stellar mass. The measurement of
$b/a$ is also the least likely to be affected by systematics from the
photometry, including the convolution of galaxy SDSS and UKIDSS images
to a common psf since it is a measure of the axial ratio in the outer
disk. 

\subsection{Physical Drivers of Residual Correlations}
\label{sec:residphys}

\begin{figure*}
\epsscale{1.0}
%\plotone{/srv/one/keckert/papers/govers/figs3/calfigs/resid2d3p.u-J.fit.new2.eps} 
\plotone{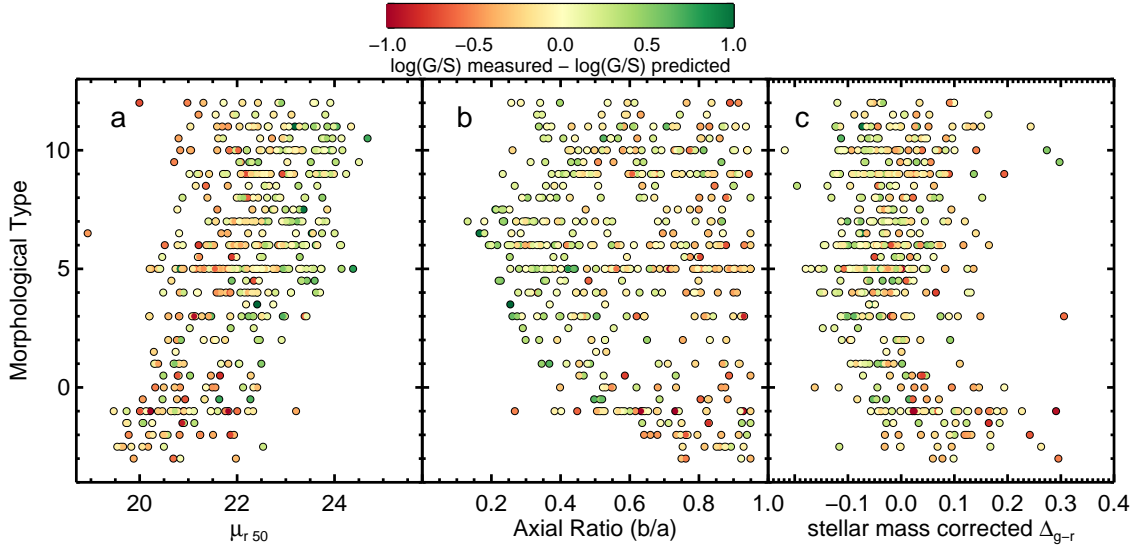} 

\caption{Residuals in log(G/S) (dot color) shown within plots of
  galaxy morphological type vs.\ (a) $\mu_{r,50}$, (b) $b/a$, and (c)
  stellar mass corrected $\Delta_{g-r}$ (following
  \citealp{2013ApJ...769...82S}).  The color scale is shown at the
  top, where green marks galaxies for which the PGF calibration
  underpredicts values of log(G/S), yellow marks galaxies for which
  the PGF calibration values of log(G/S) similar to the measured
  values, and red marks galaxies for which the PGF calibration
  overpredicts values of log(G/S). Late-type galaxies (T $>$ 0) with
  underpredicted values of log(G/S) tend to be lower surface
  brightness, more edge-on, and have redder centers (smaller values of
  stellar mass corrected $\Delta_{g-r}$). These trends become somewhat
  less noticeable for dwarf types (T $>$ 8), and in general for
  early-type galaxies values of log(G/S) tend to be overpredicted
  (with the caveat that most red-sequence early-types are redder than
  our red color cut so not shown). We discuss the physical
  significance of these patterns in \S \ref{sec:residphys}.}
\label{fg:2dresids}
\end{figure*}

For the purpose of defining PGF calibrations, we do not strictly need
to understand the drivers of the correlations, but we explore them
briefly here, deferring a more thorough discussion to future work. To
aid our interpretation, we analyze log(G/S) residuals as a function of
the best third parameter options and galaxy morphology in Figure
\ref{fg:2dresids}. Morphologies come from visual classification for
the RESOLVE data set (Moffett et al., submitted, Kannappan et al.\ in
prep.). Since galaxy morphology is generally not available for large
data sets and can be subjective and unreliable especially for small
and edge-on galaxies, we have not used morphology for the general
analysis of third parameters for improved gas mass estimation. In
Figure \ref{fg:2dresids} we plot galaxy morphology vs.\ $\mu_{r,50}$,
$b/a$, and $\Delta_{g-r}$ and color code the residuals in log(G/S)
such that green represents galaxies for which the PGF calibration
underpredicts values of log(G/S), yellow represents galaxies for which
the PGF calibration predicts values of log(G/S) similar to the
measured values, and red represents galaxies for which the PGF
calibration overpredicts values of log(G/S).

First we examine the relationship between morphological type and
$\mu_{r,50}$ in Figure \ref{fg:2dresids}a.  It appears that among
late-type galaxies (T$>$0), the PGF calibration overpredicts values of
log(G/S) for those galaxies with $\mu_{r,50}$ brighter than 22 and
underpredicts log(G/S) for those galaxies with $\mu_{r,50}$ fainter
than 22. For early-type galaxies, which mostly have $\mu_{r,50}$
brighter than 22, we generally overpredict log(G/S). The transition
around \mbox{$\mu_{r,50}$ $\sim$ 22} seems to indicate that even for a
specific type of galaxy, those galaxies that are lower surface
brightness are on average more gas-rich than their higher surface
brightness counterparts. Gas richness has long been associated with
low surface brightness galaxies (e.g.,\citealp{1996MNRAS.283...18D}
and \citealp{2004A&A...428..823O}), and we show this result for a
statistical population of galaxies using RESOLVE-A.

Next we examine the relationship between morphological type and $b/a$
in Figure \ref{fg:2dresids}b. Within the population of late-type
spirals (0 $<$ T $<$ 8), edge-on spirals tend to have underpredicted
values of log(G/S). This phenomenon makes sense for larger edge-on
spirals, which are observed to be redder than their intrinsic colors
due to dust extinction (e.g., \citealp{2010ApJ...718..184C} show that
for star forming galaxies over the narrow stellar mass range 9.5 $<$
log M$_{star}$ $<$ 10, those with $b/a$ $\sim$ 0.35 are redder than
those with $b/a$ $\sim$ 0.95). Thus the underprediction may result
because these edge-on late type spirals are shifted redward of their
true color measurement, where the typical value of log(G/S) is
smaller.\footnote{The reader may be concerned that a bias in stellar
  mass estimation as a function of extinction could lead to a spurious
  trend with axial ratio, but we have shown that our stellar masses
  are unbiased with respect to axial ratio in \S \ref{sec:mstar}.} The
trend towards underpredicting edge-on galaxy values is less noticeable
for edge-on irregular dwarf-type galaxies (T $>$ 8) that are likely
pure disk, low surface brightness galaxies with little internal
extinction due to dust (e.g., \citealp{2004ApJ...608..189D}). For
early-type galaxies, we see that in general values of log(G/S) are
overpredicted with no obvious trend with $b/a$.

\begin{figure*}
\epsscale{1.0}
%\plotone{/srv/one/keckert/papers/govers/figs3/calfigs/cal.testdust.eps} 
\plotone{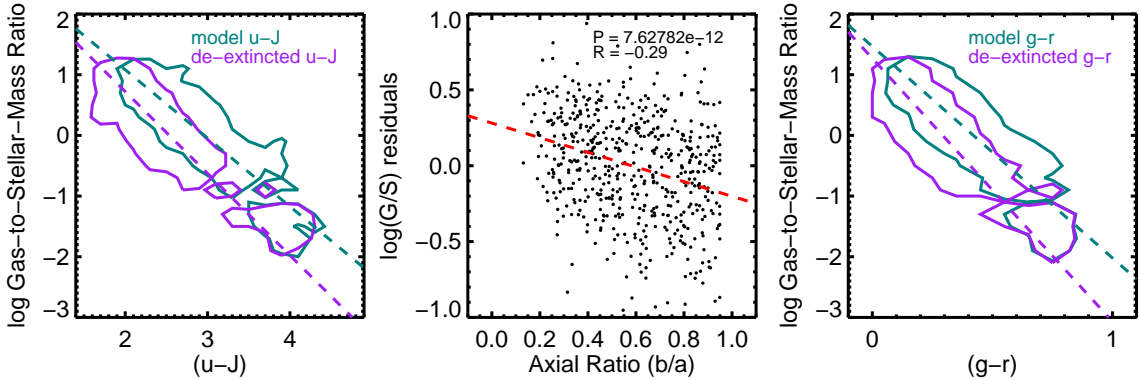}

\caption{Test of the effect of dust on the relationships between
  log(G/S), color, and axial ratio. a) Log(G/S) vs.\ \cuj{} color
  comparing the model (green) vs.\ de-extincted (purple)
  colors. Contours are shown encompassing 90\% of all galaxies
  excluding confused galaxies and weak upper limits. Dashed lines show
  the color-limited linear fits to the data. The relationship steepens
  when using the de-extincted colors. b) Versus $b/a$, the residuals
  in log(G/S) from the de-extincted \cuj{} PGF calibration have a
  smaller strength of correlation and significance than the residuals
  computed from the model \cuj{} PGF calibration, but the correlation
  is still highly significant (the red-dashed line shows the linear
  fit). c) Same as panel a except using \cgr{} color (as in
  \citealp{2015arXiv150605081J}). Unlike \citet{2015arXiv150605081J}
  we do not find that the relationship between log(G/S) and \cgr{}
  disappears completely when using de-extincted colors, although we do
  see that the relationship steepens, meaning dust is a contributing
  factor to the relationship.}
\label{fg:testdust}
\end{figure*}

We show the PGF calibrations using both model (green) and de-extincted
(purple) \cuj{} colors in Figure \ref{fg:testdust}a (we also show the
PGF calibrations using \cgr{} color in Figure \ref{fg:testdust}c). The
relationship between log(G/S) and color becomes steeper (less
predictive) when using the de-extincted rather than model colors
(Figure \ref{fg:testdust}a/c), similar to the result shown in
\citet{2015arXiv150605081J}. We do not find, however, that that the
PGF correlation completely disappears when using de-extincted colors,
suggesting that while dust plays a role in the PGF calibration,
long-term star formation is also important (K13). We suspect that star
formation history is in fact related to dust extinction, which we plan
to follow up in future work. In Figure \ref{fg:testdust}b we show the
\textit{residuals} in log(G/S) from the de-extincted PGF correlation
as a function of axial ratio, and we find that the strength and
significance of the third parameter correlation are smaller but still
highly significant. Additional effects, perhaps related to the
morphology-surface brightness correlation, may also be important.

Nevertheless, as the goal of this work is to provide empirical PGF
calibrations for predicting gas masses, we prefer not to correct for
extinction, since uncorrected colors provide the best (most linear and
most predictive) calibrations. Moreover, uncorrected colors are not as
dependent on modeling.

%Califa paper reference

%For
%the lowest $\mu_{\Delta}$ objects ($\mu_{\Delta}$ $<$ 7.5), which are
%likely pure disk, low surface brightness galaxies, there is very
%little trend with G/S residuals and axial ratio.  These dwarf galaxies
%are not likely to suffer significantly from internal extinction due to
%dust (e.g., \citealp{1999AJ....118.2751M} find little internal
%extinction in the detailed study of a super-thin disk in UGC7351).
%Climbing to higher values of $\mu_{\Delta}$ (more bulge dominated
%galaxies), we see a trend between higher G/S measured than predicted
%5for more inclined galaxies.  As the bulge becomes more prominent and
%dust a more significant part of the galaxy ISM, edge-on disks appear
%redder than their intrinsic color.  Thus these edge-on galaxies have
%higher gas content than would be predicted for their apparent color.
%If they were moved towards bluer colors, they might line up with the
%true calibration.

Lastly we examine the relationship between morphological type and
stellar mass corrected $\Delta_{g-r}$ in Figure \ref{fg:2dresids}c.
We correct $\Delta_{g-r}$ values for the correlation between
$\Delta_{g-r}$ and stellar mass as in \citet{2013ApJ...769...82S} to
show the effects more strongly. Late-type galaxies are generally more
red-centered with a slight trend towards bluer centers for later types
(larger values of T), while early-type galaxies tend to be more blue
centered as most of these are blue-sequence E/S0s
\citep{2009AJ....138..579K} because normal red-sequence E/S0s are
excluded by our color cut.

Within the late-type population, for a given morphological
type more blue centered galaxies have overpredicted values of
log(G/S). This result is intriguing because among late-type galaxies,
blue centered galaxies are associated with recent star formation
events due to interactions \citep{2004AJ....127.1371K}, and have
higher M$_{H_2}$/M$_{HI}$ ratios and lower M$_{HI}$/M$_{star}$ ratios
\citep{2013ApJ...769...82S}. Including the molecular gas for such
blue-centered late-type galaxies that are offset low in the
color-limited PGF calibration may bring their total gas-to-stellar
mass ratio (where total gas mass = 1.4M$_{HI}$ + M$_{H_2}$) in line
with the general relationship between log(G/S) and color as shown in
Figure 8 of K13.

%For galaxies with $\mu_{\Delta}$ $<$ 8.0, we find
%little trend with $g-r$ color gradient.  Above $\mu_{\Delta}$ = 8.0,
%however, we do find a trend that more blue centered galaxies (larger
%values of $g-r$ color gradient) are associated with lower G/S
%residuals.  This result is intriguing because among late-type
%galaxies, blue centered galaxies are associated with recent star
%formation events due to interactions \citep{2004AJ....127.1371K}, and
%follow a trend toward higher M$_{H2}$/M$_{HI}$ ratios and lower
%M$_{HI}$/M$_{star}$ ratios \citep{2013ApJ...769...82S}. Including the
%molecular gas for such late type galaxies that typically hang low in
%the photometric gas fractions calibration may bring total
%gas-to-stellar mass ratio (where total gas mass = 1.4$\times$M$_{HI}$
%+ M$_{H2}$) in line with the general calibration as shown in Figure 8
%of K13.

The analysis of log(G/S) residuals as a function of galaxy morphology
and the aforementioned three photometric parameters reveals physical
trends for all three that may have implications for galaxy evolution.
The two parameters $\mu_{r,50}$ and $\Delta_{g-r}$, however, are both
covariant with stellar mass, so their ability to reduce scatter is
partially artificial. Using $b/a$ allows us to minimize scatter in a
physically meaningful way without introducing covariance into the
PGF calibration.

\subsection{Modified Color-Limited PGF Calibration}
\label{sec:modcolpgfcal}

To use $b/a$ in PGF calibrations, we compute a linear combination of
color and $b/a$ to be the new predictor. We call this combination of
color and $b/a$ ``modified color.'' First, we fit a plane in color,
$b/a$, and log(G/S), minimizing scatter in log(G/S) and only using
those data points that fall within the blue color trim and red color
cutoff for a particular color choice. The planar fit determines
coefficients such that modified color = $m_0$$\times$color +
$m_1$$\times$($b/a$).

We then use the relationship between modified color and log(G/S) to
create tighter modified color-limited PGF calibrations. We use the
same procedure as in \S \ref{sec:simplecals}. First we limit our data
set to the 744 galaxies with reliable HI data. Then we define a red
modified color cutoff and a blue modified color trim just as in \S
\ref{sec:simplecals}. Lastly we perform an ordinary least squares
forward fit. Again we choose not to weight the fit by the errors on
log(G/S) due to correlations between the uncertainties and both galaxy
color and log(G/S) itself. Table \ref{tb:modcolorcals} gives the
modified color coefficients $m_0$ and $m_1$, the slope of the linear
fit, the offset in log(G/S), and the $\sigma$ of the modified color
PGF calibrations. The blue modified color trim and red modified color
cutoff are also given as well as the number of galaxies used in each
modified color fit. We note that the scatter for the PGF calibrations
using modified colors based on $u$ and $g$ are typically reduced by
$\sim$0.03 dex as compared to the PGF calibrations using color
only. Another advantage of using the modified color calibrations is
the generally increased baseline (or range) of the predictor value.

We note that many other works have used planes defined by color and
other galaxy structural parameters to define tighter PGF calibrations
\citep{2009MNRAS.397.1243Z,2010MNRAS.403..683C,2012MNRAS.424.1471L,2013MNRAS.436...34C},
often choosing stellar mass surface density or surface brightness,
which reduce scatter partially due to covariance with log(G/S). In
\citet{2013MNRAS.436...34C}, the authors use the term ``gas fraction
planes'' to refer to these PGF calibrations. We choose to use the
terminology ``modified color'' to refer specifically to the predicting
quantity, which is a linear combination of color and a structural
parameter, in this work chosen to be $b/a$.

\begin{deluxetable*}{lcccccc}
\tablecaption{Modified Color-limited Photometric Gas Fraction Calibrations}
\tablehead{\colhead{modified color} & \colhead{slope} & \colhead{log(G/S) offset} & \colhead{$\sigma$} & \colhead{blue trim} & \colhead{red cutoff} & \colhead{N galaxies}\\ \colhead{(mag)} & \colhead{(dex/mag)} & \colhead{(dex)} & \colhead{(dex)} & \colhead{(mag)} & \colhead{(mag)} &\colhead{}} 
\startdata
1.822$(u-r)^m$  + 0.609($b/a$) & -0.958 & 3.038 & 0.296 & 2.0 & 4.1 & 580 \\ 
1.454$(u-i)^m$  + 0.595($b/a$) & -0.989 & 2.887 & 0.291 & 1.8 & 3.8 & 579 \\ 
1.140$(u-J)^m$  + 0.594($b/a$) & -0.981 & 3.659 & 0.304 & 2.5 & 4.5 & 581 \\ 
1.075$(u-K)^m$  + 0.605($b/a$) & -0.952 & 4.228 & 0.302 & 3.3 & 5.0 & 525 \\ 
3.563$(g-r)^m$  + 0.534($b/a$) & -1.002 & 1.813 & 0.281 & 0.7 & 2.6 & 582 \\ 
2.444$(g-i)^m$  + 0.550($b/a$) & -0.984 & 1.881 & 0.287 & 0.8 & 2.6 & 568 \\ 
1.592$(g-J)^m$  + 0.550($b/a$) & -0.980 & 3.218 & 0.313 & 2.1 & 3.8 & 540 \\ 
1.435$(g-K)^m$  + 0.618($b/a$) & -0.864 & 3.678 & 0.306 & 2.8 & 4.5 & 488 \\ 
\enddata
\label{tb:modcolorcals}
\end{deluxetable*}

\section{Probability Density Field PGF Calibrations}
\label{sec:cals}

Due to the limitations of the color-limited linear fits, we present in
this section a new PGF calibration method that improves over the
color-limited fits in three ways.  First, the new method provides a
full probability distribution in log(G/S) for each galaxy given its
color or modified color rather than a single number for log(G/S).
Second, we are able to model the galaxy distribution past the red
color cutoff region where quenched galaxies coexist with star-forming
galaxies. Third, in the spirit of survival analysis, we develop a
Monte Carlo method to reinsert the 22\% of RESOLVE-A galaxies
with unreliable HI data in the calibration, thus determining the
calibration for the entire volume-limited data set. We describe the
full probability distribution method in \S \ref{sec:caldesc}, and we
outline the Monte Carlo method to reinsert unreliable galaxies in \S
\ref{sec:reinsert}. Instructions on how to use the calibration are
provided in \S \ref{sec:howtouse}. Lastly in \S \ref{sec:addhmlgals},
we examine whether the model changes significantly if we add in the
small subset of galaxies that have M$_{r,tot}$ fainter than $-17.33$,
but still have M$_{bary}$ $>$ 10$^{9.3}$ \msun. For simplicity,
throughout these sections we use the variable ``$mc$'' to refer to
either color alone or modified color that includes a $b/a$ term.

\subsection{Calibration Description}
\label{sec:caldesc}

\begin{figure*}
\epsscale{1.}
%\plotone{/srv/one/keckert/papers/govers/figs3/calfigs/goversoneuminusJ_ba.eps}
\plotone{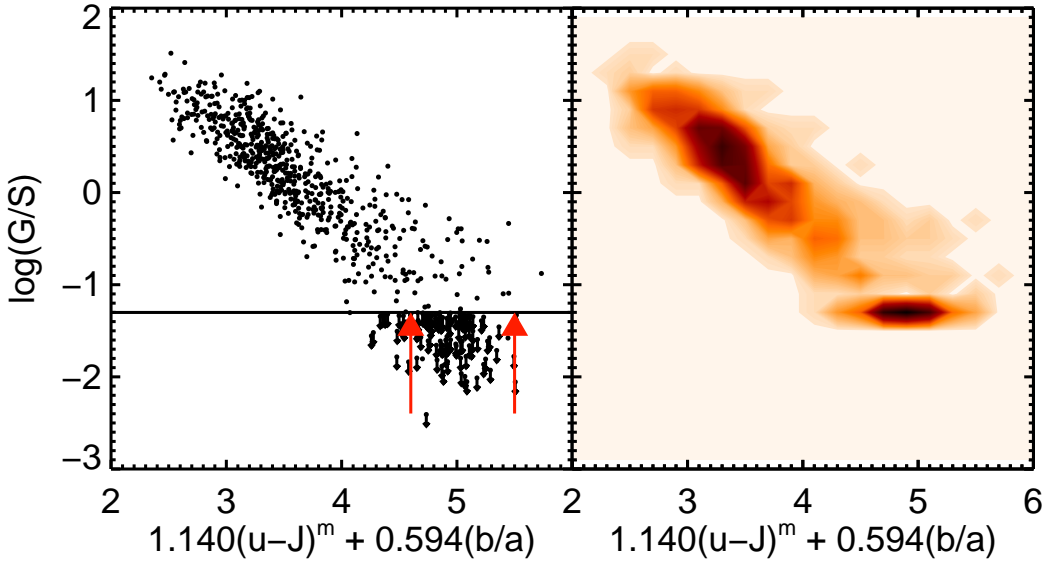}
\caption{Log(G/S) vs.\ modified \cujm{} color for RESOLVE-A. The
  modified color is the linear combination of \cujm{} and $b/a$ that
  produces the best plane fit with log(G/S) as described in \S
  \ref{sec:modcolpgfcal}. a) Scatter plot showing reliable HI
  detections with small black dots, and strong upper limits with
  downward black arrows. Galaxies with log(G/S) values $<$ $-1.3$
  (marked by the black line), whether they are detections or limits, are
  replaced with $-1.3$ (as indicated by the red arrows). b)
  Contour plot showing the PGF density field of log(G/S) vs.\ modified
  \cujm{} color, where each cell's value is the number of galaxies in
  that cell. The contours are spaced finely over 100 intervals between
  0 and 34 (the maximum number of galaxies found in a cell). We fit the
  model described in \S \ref{sec:caldesc} to this PGF density field.}

\label{fg:goversone}
\end{figure*}

To start we use the same RESOLVE-A galaxies as in \S
\ref{sec:simplecals}, those with reliable detections or strong upper
limits.  Detections are considered reliable if they are not confused
or are deconfused with systematic errors $<$ 25\% of the HI mass
measurement and have S/N $>$ 5. Limits are considered strong if they
provide a gas mass measurement of M$_{gas}$ $<$ 0.05M$_{star}$. We
show log(G/S) vs.\ modified \cujm{} color for the RESOLVE-A data set
limited to reliable gas data in Figure \ref{fg:goversone}a. This
definition excludes 211 galaxies (of 955 total) in RESOLVE-A, however
we describe how a technique based on survival analysis principles to
include these galaxies in the PGF calibration in \S
\ref{sec:reinsert}.

For galaxies with upper limits, we cannot know the value of the HI
mass, only that it is less than a certain value. We choose to set all
galaxies, both detections and limits, with log(G/S) $<$ $-1.3$ to
log(G/S) = $-1.3$, which is 5\% of the stellar mass in linear
units. This choice is fine for measuring a galaxy's baryonic mass (as
in the companion paper Eckert et al.\ in prep.), because a galaxy's
baryonic mass will not be significantly affected if it is at most
1.05M$_{star}$. In Figure \ref{fg:goversone}a, we demonstrate the data
replacement with red arrows.

Next we determine the density field of galaxies in log(G/S) vs.\ $mc$
by binning in both dimensions and measuring the number of galaxies in
each 2D bin or cell of log(G/S) and $mc$.  The density field is shown
in Figure \ref{fg:goversone}b.  We use bin sizes of 0.2 dex in
log(G/S) and 0.2 mag in \cujm{}.  When determining the bin size for
each $mc$, we start with bin sizes of 0.2 mag and require a minimum of
10 bins in color to ensure adequate sampling of the calibration. If
there are less than 10 bins (as is the case for colors with small
ranges such as \cgr) we switch to bin sizes of 0.1 mag.

To model the PGF probability density field we assume two distinct
populations: detections and upper limits.  The detections lie on a
line with Gaussian scatter that widens towards redder colors.  The
limits appear only at red $mc$ and are confined to a value of log(G/S)
= $-1.3$ although they may in fact have much lower values. The model
consists of nine parameters given by $A_0$ to $A_8$.

% In \S
%\ref{sec:detections} we describe the model for the detections and in
%\S \ref{sec:limits} we describe the model for the limits. The model
%consists of nine parameters given by A$_0$ through A$_8$.

%\subsubsection{Detections}
%\label{sec:detections}

To model the population of detections we first model the peak value of
the Gaussian at each $mc$, designated as $\rho_0$, as a log-normal
function of $mc$ as given in equation \ref{eq:peak}. $A_0$ provides
the normalization, $A_1$ is the location parameter in $mc$, and $A_2$
is the shape parameter.

\begin{equation}
\rho_0 = \frac{A_0}{(mcA_2\sqrt{2\pi})}\times\exp(\frac{-(\ln(mc)-A_1)^2}{2A_2^2})
\label{eq:peak}
\end{equation}

The number density of detections ($D$) is then described by a Gaussian
with peak value $\rho_0$ that changes with $mc$, with mean log(G/S)
value ($A_3mc + A_4$) that decreases linearly for larger values of
$mc$, and with standard deviation $A_5mc$ that widens towards larger
values of $mc$. The model for the detections is given in equation
\ref{eq:z1a}.

\begin{equation}
D =\rho_0\times\exp(\frac{-(log(G/S)-(A_3mc+A_4))^2}{2(A_5mc)^2})
\label{eq:z1a}
\end{equation}

At red $mc$, there are galaxies with values of \mbox{log(G/S) $<$
  $-1.3$} that belong to the detection rather than limit
population. Because we have chosen to set all galaxies with values of
log(G/S) $<$ $-1.3$ to be equal to $-1.3$, we must divide up equation
\ref{eq:z1a} to account for the tail of the detection population with
log(G/S) $<$ $-1.3$. For values of log(G/S) $>$ $-1.3$, the model is
as given by equation \ref{eq:z1a}, but for values of log(G/S) =
$-1.3$, we model the detection population by integrating the tail of
the Gaussian at a given $mc$ over all log(G/S) $<$ $-1.3$. Thus the
detection population for log(G/S) $<$ $-1.3$ is given in equation
\ref{eq:z1b}.

\begin{equation}
%\begin{split}
D=\rho_0\frac{1}{2}A_5mc\sqrt{2\pi}\times[1-erf(\frac{\left |-1.3-(A_3mc+A_4)  \right |}{\sqrt{2(A_5mc)^2}})]
%\end{split}
\label{eq:z1b}
\end{equation}

The upper limit population is evaluated only at \mbox{log(G/S) $=$ $-1.3$}
and is modeled as a Gaussian function of $mc$ with a peak value $A_6$,
color shift $A_7$, and standard deviation $A_8$ as given in equation
\ref{eq:z2}.  The Gaussian shape captures the fact that blue galaxies
rarely have low gas fractions and on the other side that there is a
natural decline in galaxies towards redder colors.

\begin{equation}
L= A_6\times\exp(\frac{-(mc-A_7)^2}{2A_8^2})%\times(c = -1.3)
\label{eq:z2}
\end{equation}

The full-probability PGF model is the combination of equations
\ref{eq:z1a}, \ref{eq:z1b}, and \ref{eq:z2}, and we fit this model to
the PGF density field using the MPFIT2DFUN package
\citep{2009ASPC..411..251M}\footnote{http://purl.com/net/mpfit}, which
performs a Levenberg-Marquardt least squares fit \citep{more}. We
weight each cell value in the density field, which is the number of
galaxies in the cell N, by 1/N. From the fit we obtain the parameters
$A_{0}$ through $A_{8}$ that best describe the PGF calibration. We
recognize that the probability density field model is more complex
than a simple linear fit. However, the advantages of being able to fit
both the detection and limit populations and of being able to reinsert
the data from galaxies without adequate HI data (see \S
\ref{sec:reinsert}) make it a more powerful technique for estimating
gas masses (see \S \ref{sec:discussion}).

\subsection{Reinserting the Missing Galaxies}
\label{sec:reinsert}

To measure probability density field PGF calibrations with the
complete RESOLVE-A data set, we implement a custom version of survival
analysis following its basic algorithm: create a model of the
missing/censored data points based closely on the available data and
sample it with Monte Carlo methods. The model for each
censored/missing data point is determined from the uncensored data in
an initial fitting round as in traditional survival analysis, with the
restriction that limits must be drawn from a model distribution
truncated at the limit values. We then iterate this procedure a second
time using the updated model resulting from the first round. The steps
are as follows.

1) After performing the initial PGF model fit from \S
\ref{sec:caldesc} using only the reliable HI data in RESOLVE-A, we
obtain a PGF model with initial best fit parameters $A_{i,0}$ through
$A_{i,8}$. Using these initial best fit parameters, we obtain a
probability distribution in log(G/S) for each galaxy with unreliable
HI data given its $mc$. For galaxies with weak upper limits, we
restrict the probability distributions in log(G/S) to only those
values below the upper limit, then we renormalize the probability
distribution.  For low S/N detections and confused galaxies we do not
place any restrictions on the probability distribution in log(G/S).

2) We then randomly assign a value of log(G/S) for each of these
211 galaxies from each respective probability
distribution using the inverse transform sampling method.

3) Next, we create a new PGF density field that includes both the 744
data points with reliable HI and the 211 data points with randomly
assigned values of log(G/S) from their respective probability
distributions. We fit the PGF calibration model as described in \S
\ref{sec:caldesc} and save the best fit parameters $A_{n,0}$ through
$A_{n,8}$. We also compute residuals for each best fit model as well
as the reduced $\chi^2$ goodness of fit, assuming the variance on the
data value for each cell is $\sigma^2$ = N (the number of galaxies in
the cell).

4) We perform steps 2 \& 3 100 times and save the best fit parameters
and reduced $\chi^2$ values for each round. We then calculate the
median values of the parameters based on those 100 rounds $A_{med,0}$
through $A_{med,8}$, which describe the new model for the PGF density
field. We save the median reduced $\chi^2$ as well.

5) We iterate once more through steps 1-4 using the model parameters
$A_{med,0}$ through $A_{med,8}$ to produce the initial probability
distributions in log(G/S) for the 211 galaxies with unreliable HI
data. This iteration allows us to determine log(G/S) probability
distributions based on the PGF density field calibrated with all
RESOLVE-A galaxies, not just those with reliable HI data. After this
second iteration, we take the values of $A_{med,0}$ through
$A_{med,8}$ to be the final best fit parameters $A_{f,0}$ through
$A_{f,8}$ for the full-probability PGF model. We have tested whether
more iterations are necessary, however after the second iteration the
values of $A_{med,0}$ through $A_{med,8}$ change very little ($<$ 3\%
of the final value). The final values do change significantly from the
initial fit (which includes only the reliable HI galaxies), with final
values up to $\sim$20\% different from the initial values.

\begin{figure*}
\epsscale{1.}
%\plotone{/srv/one/keckert/papers/govers/figs3/calfigs/goversmodelandresidsuminusJ_ba.eps}
\plotone{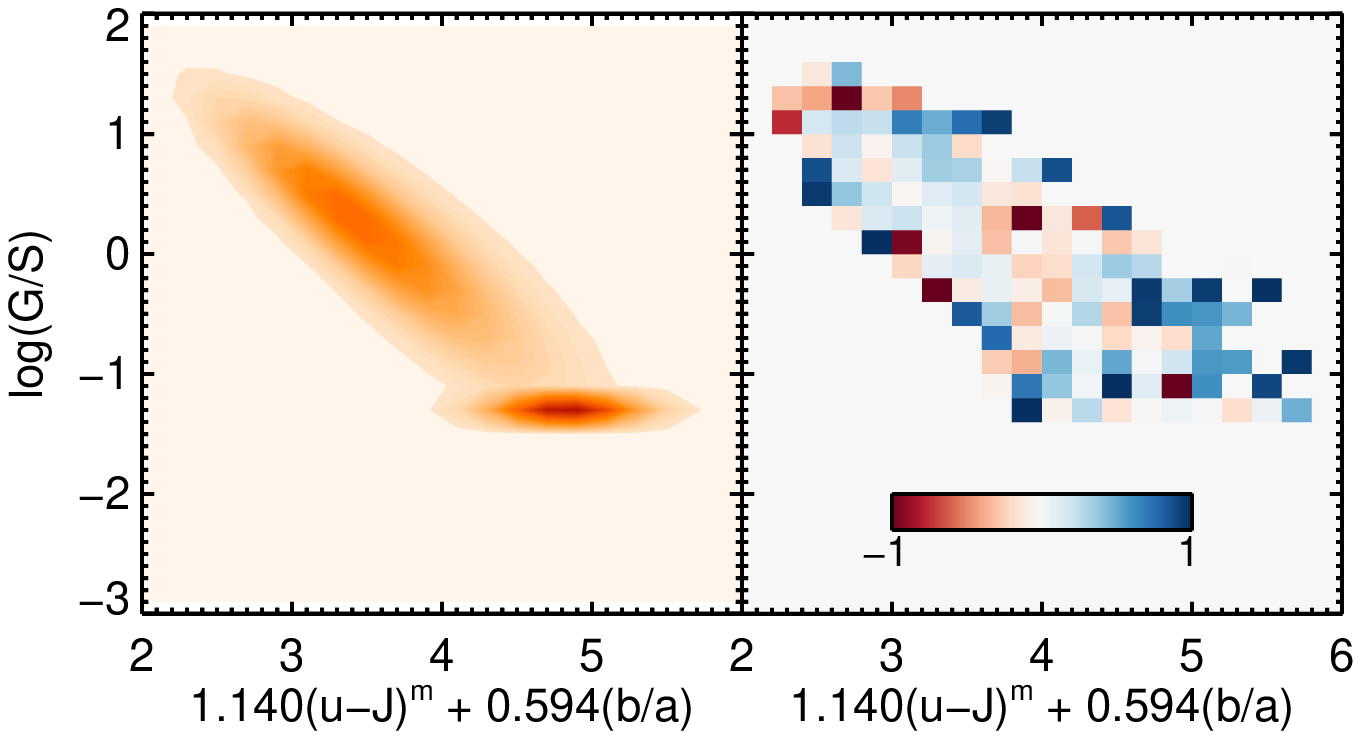}
\caption{Best fit model to the PGF density field using modified
  \cujm{} color.  a) The best fit model is shown as a contour plot (as
  for Figure \ref{fg:goversone}b divided into 100 levels between 0 and
  44, the maximum number of galaxies found in a cell after reinsertion
  of galaxies with inadequate HI data). b) Residuals for each cell
  normalized by the number of galaxies in each cell so the range is
  from $-1$ to 1.  Overall the model performs well, although there are
  large residuals especially around the edges of the calibration where
  the number density of galaxies is low.}

\label{fg:goverstwo}
\end{figure*}

Figure \ref{fg:goverstwo} shows a contour plot of the best fit model
and residuals using modified \cujm{} color, where the residuals have
been normalized by the number of galaxies in each cell. The 211
galaxies with unreliable HI data have been assigned values of log(G/S)
pulled at random from their respective distributions in log(G/S), as
in step 2, using the final best fit parameters. The residuals are most
significant for the highest log(G/S) cells at red modified color where
there are few galaxies (often just one per cell).

\subsection{How to Use the PGF Calibration}
\label{sec:howtouse}

\begin{figure}
\epsscale{1.}
%\plotone{/srv/one/keckert/papers/govers/figs3/calfigs/probsuminusJ_ba.eps}
\plotone{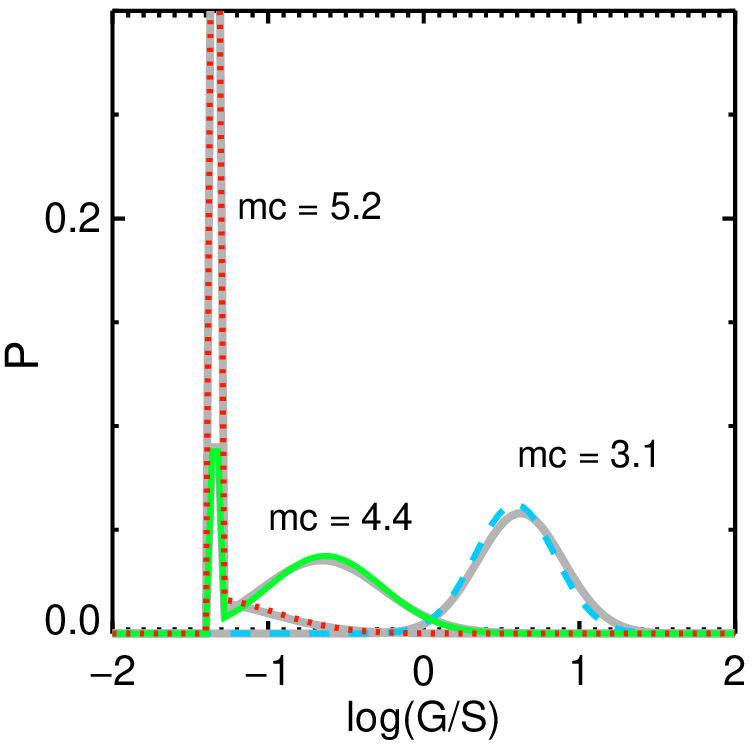}
\caption{Probability distributions for three galaxies with modified
  \cuj$^m$ ``$mc$'' = 3.1, 4.4, and 5.2, which are labeled on the
  plot. For blue modified colors the distribution in log(G/S) is
  Gaussian ($mc$ = 3.1). The $\sigma$ value of the Gaussian widens for
  redder modified colors ($mc$ = 4.4), where the population of
  galaxies with log(G/S) $<$ $-1.3$ also increases. The sharp spike at
  log(G/S) = $-1.3$ is due to the way in which galaxies with upper
  limits or low gas-fractions are treated in the model. For the
  reddest colors ($mc$ = 5.2)the contribution from the upper limit
  population with log(G/S) $<$ $-1.3$ becomes dominant. In solid grey
  we show the probability distributions for the same modified colors
  when using the fits that include high baryonic mass-to-light
  galaxies as explained in \ref{sec:addhmlgals}. The largest
  difference occurs for the bluest galaxy, although the difference is
  quite small.}

\label{fg:goversprob}
\end{figure}

The parameters in Tables \ref{tb:newparamscoloronly} and
\ref{tb:newparamscolorandba} provide the best fit models to the PGF
density field probability distribution, which is not normalized at a
given value of $mc$. To create conditional probability distributions
in log(G/S) at each $mc$, the PGF distribution for a given $mc$ from
equations \ref{eq:z1a}, \ref{eq:z1b}, and \ref{eq:z2} should be
divided by the integral of the PGF at that $mc$.\footnote{We provide
  an IDL code at
  https://github.com/keckert7/codes/pred\_loggs\_dist.pro that looks
  up the best fit parameters for a given $mc$ and provide the
  probability distributions for any set of galaxies with provided
  $mc$.} Examples of such distributions are shown in Figure
\ref{fg:goversprob} for modified \cujm{} color $mc$ = 3.1 (blue), 4.4
(green), and 5.3 (red).  We note that the sharp spike in the log(G/S)
probability distribution is entirely due to the way in which upper
limits and low gas-fraction galaxies are treated in the model and
should not be interpreted literally as probability distributions for
this population.

%/$\sim$keckert/code/pred_loggs_dist.pro

%Once we have the 2D functional fit, we can compute an array of
%probabilities for each value of G/S ranging from G/S = -1.3 to 2 for
%any galaxy with a given color or color plus surface brightness. Since the
%function is given by the best fit parameters to the full data, we must
%afterwards normalize the probability distribution such that the sum of
%the probabilities for a given galaxy with some color is one.  If we
%want to determine a particular G/S value for the galaxy, we compute
%the median value of G/S for the probability distribution.  In other
%applications, such as creating the baryonic mass function (Eckert et
%al., in prep) we may choose to use the entire distribution of G/S to
%fully model the gas distributions of our galaxy population.

%\input{/srv/one/keckert/papers/govers/figs3/calfigs/2Dparamfitstablecoloronly_072415.tex}
\begin{deluxetable*}{ccccccccccc}
\tablecaption{Full-Probability Photometric Gas Fractions Calibrations for Color Only}
\tablehead{\colhead{color} & \colhead{$A_{f,0}$} & \colhead{$A_{f,1}$} & \colhead{$A_{f,2}$} & \colhead{$A_{f,3}$} & \colhead{$A_{f,4}$} & \colhead{$A_{f,5}$} & \colhead{$A_{f,6}$} & \colhead{$A_{f,7}$} & \colhead{$A_{f,8}$} & \colhead{reduced $\chi^2$}\\ \colhead{(mag)} & \colhead{} & \colhead{} & \colhead{} & \colhead{} & \colhead{} & \colhead{} & \colhead{}& \colhead{} & \colhead{} & \colhead{}} 
\startdata
$(u-r)^m$ & 17.22 & 0.33 & 0.22 & -1.67  & 2.56  & 0.21  & 47.65  & 2.21  & 0.14 & 1.353 \\ 
$(u-i)^m$ & 35.05 & 0.46 & 0.24 & -1.37  & 2.40  & 0.19  & 68.44  & 2.55  & 0.20 & 1.518 \\ 
$(u-J)^m$ & 32.55 & 1.00 & 0.17 & -1.04  & 3.07  & 0.11  & 53.39  & 3.88  & 0.26 & 1.257 \\ 
$(u-K)^m$ & 30.18 & 1.26 & 0.13 & -0.99  & 3.75  & 0.09  & 42.35  & 4.79  & 0.33 & 1.422 \\ 
$(g-r)^m$ & 19.70 & -1.15 & 0.48 & -3.65  & 1.51  & 0.87  & 115.1  & 0.72  & 0.05 & 1.831 \\ 
$(g-i)^m$ & 19.99 & -0.70 & 0.45 & -2.47  & 1.58  & 0.55  & 72.29  & 1.05  & 0.09 & 1.344 \\ 
$(g-J)^m$ & 16.99 & 0.51 & 0.19 & -1.52  & 2.80  & 0.18  & 42.87  & 2.37  & 0.16 & 1.036 \\ 
$(g-K)^m$ & 31.08 & 0.91 & 0.15 & -1.22  & 3.27  & 0.14  & 61.95  & 3.29  & 0.22 & 1.246 \\ 
\enddata
\label{tb:newparamscoloronly}
\end{deluxetable*}

\begin{deluxetable*}{ccccccccccc}
\tablecaption{Full-Probability Photometric Gas Fractions Calibrations for Modified Color}
\tablehead{\colhead{modified color} & \colhead{$A_{f,0}$} & \colhead{$A_{f,1}$} & \colhead{$A_{f,2}$} & \colhead{$A_{f,3}$} & \colhead{$A_{f,4}$} & \colhead{$A_{f,5}$} & \colhead{$A_{f,6}$} & \colhead{$A_{f,7}$} & \colhead{$A_{f,8}$} & \colhead{reduced $\chi^2$}\\ \colhead{(mag)} & \colhead{} & \colhead{} & \colhead{} & \colhead{} & \colhead{} & \colhead{} & \colhead{}& \colhead{} & \colhead{} & \colhead{}} 
\startdata
1.822$(u-r)^m$ + 0.609(b/a) & 38.78 & 1.05 & 0.20 & -0.89  & 2.80  & 0.09  & 47.95  & 4.44  & 0.29 & 1.125\\ 
1.454$(u-i)^m$ + 0.595(b/a) & 37.78 & 0.97 & 0.21 & -0.93  & 2.70  & 0.10  & 51.88  & 4.10  & 0.27 & 1.104\\ 
1.140$(u-J)^m$ + 0.594(b/a) & 38.13 & 1.25 & 0.16 & -0.93  & 3.49  & 0.08  & 44.57  & 4.82  & 0.31 & 1.066\\ 
1.075$(u-K)^m$ + 0.605(b/a) & 34.54 & 1.43 & 0.14 & -0.88  & 3.92  & 0.07  & 40.53  & 5.53  & 0.34 & 1.157\\ 
3.563$(g-r)^m$ + 0.534(b/a) & 42.65 & 0.37 & 0.37 & -0.97  & 1.74  & 0.18  & 62.12  & 2.92  & 0.21 & 1.166\\ 
2.444$(g-i)^m$ + 0.550(b/a) & 40.95 & 0.45 & 0.36 & -0.99  & 1.87  & 0.17  & 60.79  & 2.95  & 0.21 & 1.228\\ 
1.592$(g-J)^m$ + 0.550(b/a) & 36.40 & 1.10 & 0.18 & -0.94  & 3.08  & 0.10  & 51.50  & 4.16  & 0.26 & 1.028\\ 
1.435$(g-K)^m$ + 0.618(b/a) & 33.79 & 1.37 & 0.14 & -0.91  & 3.83  & 0.08  & 45.48  & 5.10  & 0.31 & 1.182\\ 
\enddata
\label{tb:newparamscolorandba}
\end{deluxetable*}

\subsection{Adding in High M$_{bary}$/L Galaxies}
\label{sec:addhmlgals}

As a check, we have also performed the full-probability PGF
calibration using a slightly altered RESOLVE-A data set. This data set
aligns with the spectroscopic observing data set for the RESOLVE
survey. We consider all galaxies with M$_{r,tot}$ brighter than
$-17.33$ as well as any galaxies with \mbox{M$_{bary}$ $>$ 10$^{9.3}$
  \msun{}} no matter their M$_{r,tot}$, thus including the highest
baryonic mass-to-light ratios that fall out of the absolute magnitude
limited sample.

For this test we use the PGF calibrations from Tables
\ref{tb:newparamscoloronly} and \ref{tb:newparamscolorandba} to
estimate gas masses for the galaxies with M$_{r,tot}$ fainter than
$-17.33$ that do not have adequate gas data. Using the median value of
log(G/S) from each probability distribution we estimate the gas mass
for these galaxies. We then compute M$_{bary}$ for each galaxy in the
RESOLVE-A region, adding the stellar mass to either the measured
M$_{gas}$ for those galaxies with reliable detections and strong upper
limits, or the predicted M$_{gas}$ for those galaxies with low S/N
detections, confused profiles, or weak upper limits. Adding galaxies
with M$_{r,tot}$ fainter than $-17.33$ and M$_{bary}$ $>$ 10$^{9.3}$
\msun{} yields a RESOLVE-A data set with 39-40 additional galaxies
depending on the choice of PGF calibration. Thirty-nine of those
galaxies have reliable HI data.

With these galaxies added, we can now run through the same procedure
as detailed in \S \ref{sec:caldesc} and \S \ref{sec:reinsert}. The
parameters for the best fit models are given in Tables
\ref{tb:newparamscoloronlyhml} and
\ref{tb:newparamscolorandbahml}. The PGF calibrations yield similar
log(G/S) distributions with or without the high baryonic mass-to-light
ratio galaxies. In Figure \ref{fg:goversprob} grey solid lines show
the log(G/S) probability distributions for $mc$ = 3.1, 4.4, and 5.2 when
including the high baryonic mass-to-light ratio galaxies. The largest
difference is seen for the bluest galaxies, where the high
baryonic-mass-to-light galaxies very slightly shift the probability
distribution towards higher values of log(G/S).

\begin{deluxetable*}{ccccccccccc}
\tablecaption{Full-Probability Photometric Gas Fractions Calibrations for Color Only Including High M$_{bary}$/L Galaxies}
\tablehead{\colhead{color} & \colhead{$A_{f,0}$} & \colhead{$A_{f,1}$} & \colhead{$A_{f,2}$} & \colhead{$A_{f,3}$} & \colhead{$A_{f,4}$} & \colhead{$A_{f,5}$} & \colhead{$A_{f,6}$} & \colhead{$A_{f,7}$} & \colhead{$A_{f,8}$} & \colhead{reduced $\chi^2$}\\ \colhead{(mag)} & \colhead{} & \colhead{} & \colhead{} & \colhead{} & \colhead{} & \colhead{} & \colhead{}& \colhead{} & \colhead{} & \colhead{}} 
\startdata
$(u-r)^m$ & 34.56 & 0.31 & 0.23 & -1.66  & 2.57  & 0.24  & 90.35  & 2.21  & 0.15 & 1.514 \\ 
$(u-i)^m$ & 35.01 & 0.44 & 0.24 & -1.44  & 2.54  & 0.20  & 68.48  & 2.55  & 0.20 & 1.474 \\ 
$(u-J)^m$ & 32.37 & 1.00 & 0.17 & -1.10  & 3.28  & 0.12  & 53.83  & 3.88  & 0.26 & 1.308 \\ 
$(u-K)^m$ & 29.47 & 1.26 & 0.13 & -1.03  & 3.93  & 0.10  & 42.53  & 4.78  & 0.33 & 1.446 \\ 
$(g-r)^m$ & 19.24 & -1.17 & 0.49 & -3.74  & 1.56  & 0.93  & 115.8  & 0.72  & 0.05 & 2.033 \\ 
$(g-i)^m$ & 19.75 & -0.71 & 0.45 & -2.54  & 1.63  & 0.58  & 72.31  & 1.05  & 0.09 & 1.514 \\ 
$(g-J)^m$ & 32.44 & 0.50 & 0.20 & -1.54  & 2.87  & 0.22  & 82.06  & 2.39  & 0.17 & 1.098 \\ 
$(g-K)^m$ & 29.46 & 0.91 & 0.15 & -1.29  & 3.46  & 0.15  & 62.12  & 3.29  & 0.23 & 1.323 \\ 
\enddata
\label{tb:newparamscoloronlyhml}
\end{deluxetable*}

\begin{deluxetable*}{ccccccccccc}
\tablecaption{Full-Probability Photometric Gas Fractions Calibrations for Modified Color Including High M$_{bary}$/L Galaxies}
\tablehead{\colhead{modified color} & \colhead{$A_{f,0}$} & \colhead{$A_{f,1}$} & \colhead{$A_{f,2}$} & \colhead{$A_{f,3}$} & \colhead{$A_{f,4}$} & \colhead{$A_{f,5}$} & \colhead{$A_{f,6}$} & \colhead{$A_{f,7}$} & \colhead{$A_{f,8}$} & \colhead{reduced $\chi^2$}\\ \colhead{(mag)} & \colhead{} & \colhead{} & \colhead{} & \colhead{} & \colhead{} & \colhead{} & \colhead{}& \colhead{} & \colhead{} & \colhead{}} 
\startdata
1.822$(u-r)^m$ + 0.609(b/a) & 38.82 & 1.05 & 0.21 & -0.90  & 2.84  & 0.10  & 48.06  & 4.44  & 0.29 & 1.194\\ 
1.454$(u-i)^m$ + 0.595(b/a) & 37.44 & 0.96 & 0.21 & -0.96  & 2.79  & 0.11  & 51.32  & 4.10  & 0.27 & 1.188\\ 
1.140$(u-J)^m$ + 0.594(b/a) & 36.98 & 1.24 & 0.16 & -0.98  & 3.66  & 0.08  & 44.90  & 4.82  & 0.31 & 1.154\\ 
1.075$(u-K)^m$ + 0.605(b/a) & 33.11 & 1.43 & 0.14 & -0.91  & 4.05  & 0.07  & 40.99  & 5.53  & 0.34 & 1.219\\ 
3.563$(g-r)^m$ + 0.534(b/a) & 41.40 & 0.36 & 0.37 & -1.01  & 1.82  & 0.19  & 61.93  & 2.91  & 0.21 & 1.299\\ 
2.444$(g-i)^m$ + 0.550(b/a) & 39.53 & 0.44 & 0.35 & -1.03  & 1.97  & 0.18  & 60.35  & 2.95  & 0.21 & 1.309\\ 
1.592$(g-J)^m$ + 0.550(b/a) & 35.19 & 1.10 & 0.18 & -0.97  & 3.20  & 0.10  & 51.74  & 4.16  & 0.26 & 1.171\\ 
1.435$(g-K)^m$ + 0.618(b/a) & 32.02 & 1.37 & 0.14 & -0.94  & 3.96  & 0.08  & 45.97  & 5.10  & 0.31 & 1.216\\ 
\enddata
\label{tb:newparamscolorandbahml}
\end{deluxetable*}

\section{Discussion}
\label{sec:discussion}

In this section we perform two comparisons of PGF calibrations from
this work and other work using the RESOLVE-B data set. First, we
compare predicted gas masses from PGF calibrations with measured gas
masses from the RESOLVE-B data set as a function of stellar mass. To
test the different PGF calibrations, we select only galaxies in
RESOLVE-B that have reliable HI detections (294 out of 487), as upper
limits are not ideal candidates for this test. Second, we compare the
actual RESOLVE-B distribution of log(G/S) in bins of \cujm{} color
with the predicted distributions of log(G/S) based on these different
calibrations.

%%changing order of probability and color-limited

\begin{figure*}
\epsscale{1.}
%\plotone{/srv/one/keckert/papers/govers/figs3/finalfig.eps} 
\plotone{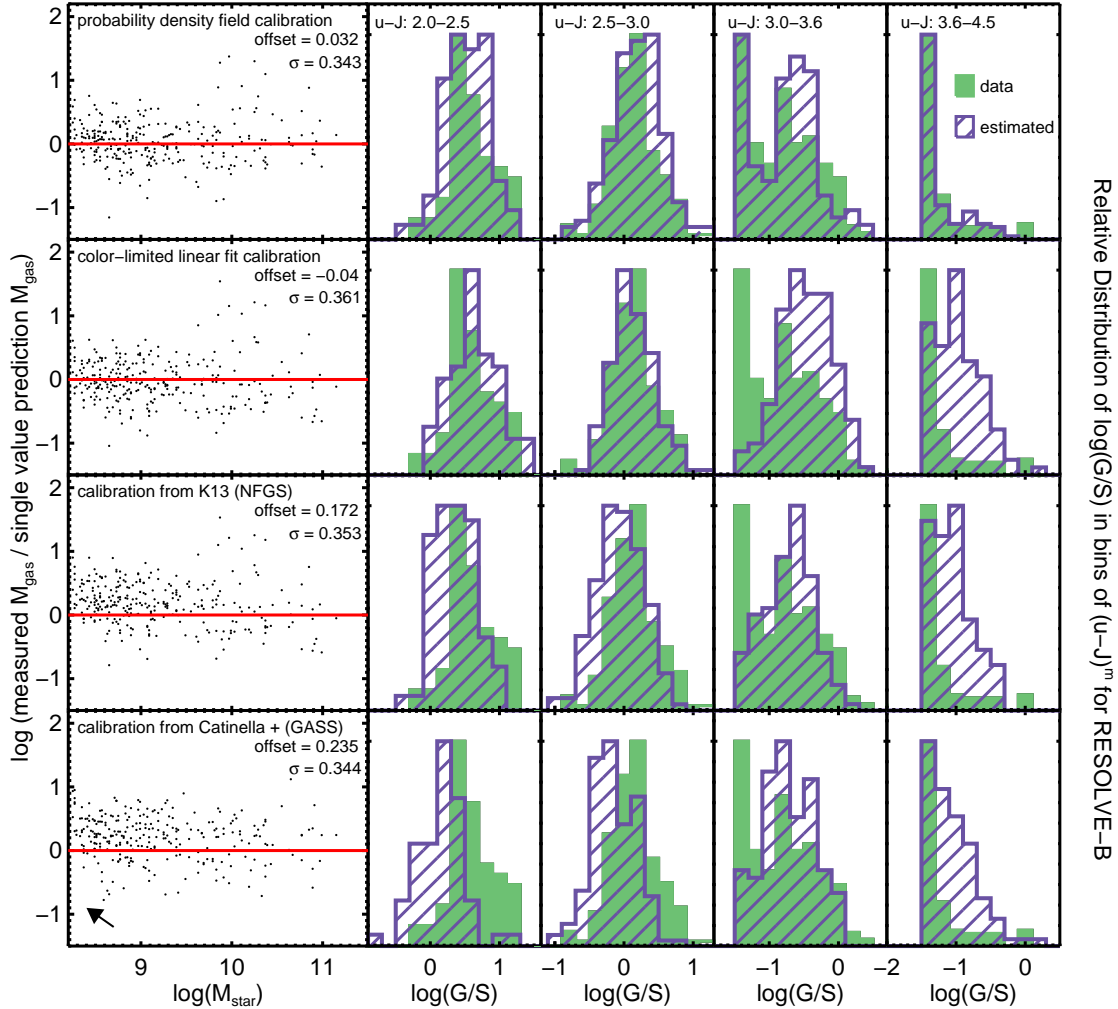} 

\caption{Comparison of PGF calibrations from this work and the
  literature using 21cm data for RESOLVE-B.  Each row shows a
  different calibration. \textbf{First Column:} Log of measured over
  predicted gas mass for RESOLVE-B galaxies with HI detections as a
  function of stellar mass. For the probability density field model
  (top row), we use the median value of each galaxy's log(G/S)
  distribution as the single value gas mass prediction, and we find a
  negligible offset and and small scatter in the residuals. The
  color-limited linear fit from this work (2nd row) also produces a
  negligible offset, but slightly larger scatter in the residuals. The
  color-limited linear fits from K13 and \citet{2013MNRAS.436...34C}
  (3rd and 4th rows) both yield large offsets towards underpredicting
  gas masses. The asymmetric scatter for low stellar mass galaxies is
  caused by the covariance between stellar mass and the estimated gas
  mass in the denominator of the y-axis (arrow in bottom
  panel). Comparisons to other calibrations
  (\citealp{2004ApJ...611L..89K}, \citealp{2009MNRAS.397.1243Z}, and
  \citealp{2012ApJ...756..113H}) are discussed in the text (\S
  \ref{sec:discussion}). \textbf{2nd-5th Columns:} Actual (green) and
  estimated (purple) distributions of log(G/S) for RESOLVE-B, where
  all values of log(G/S) $<$ $-1.3$ are set equal to $-1.3$. For the
  probability density field (top row), we draw random values from each
  galaxy's log(G/S) distribution. The estimated log(G/S) distributions
  are consistent with the data for all colors. For the color-limited
  linear fits, we add random scatter with $\sigma$ $\sim$ 0.3 to the
  log(G/S) estimates for each galaxy. The color-limited linear fit
  from this work (2nd row) performs well for blue galaxies, while the
  color-limited linear fits from K13 and \citet{2013MNRAS.436...34C}
  (3rd and 4th rows) underestimate the distributions of log(G/S) for
  blue galaxies. None of the color-limited linear fits performs well
  for red galaxies, necessitating the probability density field
  model. We attribute the underprediction of gas masses for low mass
  and blue galaxies by the K13 and \citet{2013MNRAS.436...34C}
  calibrations to their samples (NFGS and GASS respectively) not being
  volume-limited samples of the galaxy population in the nearby
  universe.}

\label{fg:testgs}
\end{figure*}

To test the probability density field model we use the modified
\cujm{} color calibration (top row of Figure \ref{fg:testgs}). Since
the calibration provides a probability distribution of log(G/S) for
each galaxy, we use the median value from the probability distribution
in log(G/S) for the purposes of the first test.  In the top panel of
Figure \ref{fg:testgs} we find that using the probability density
field model PGF calibration yields a negligible offset and small
scatter = 0.343 dex. For the lowest stellar masses, we see asymmetric
scatter towards larger measured than predicted gas masses. As
indicated by the arrow in the bottom panel of Figure \ref{fg:testgs},
this asymmetric scatter is consistent with covariance between stellar
mass, plotted on the x-axis, and the predicted gas mass in the
denominator of the y-axis (note that estimated G/S is multiplied by
stellar mass to obtain gas masses). We also show the distribution of
log(G/S) predicted for RESOLVE-B galaxies in four bins of \cujm{}
color, drawing a random value from the log(G/S) distribution of each
RESOLVE-B galaxy. The probability density field model works well
across all colors.

To test our color-limited linear fits, we use the calibration based on
\cujm{} color. While the \cujm{} calibration has higher scatter than
calibrations based on optical colors, the \cujm{} calibration has a
larger predictive range as compared to typical errors on color
(considering the slopes that would multiply the error; see Tables
\ref{tb:simplecals} and \ref{tb:modcolorcals}. In the second row of
Figure \ref{fg:testgs}, we show the log of the measured gas mass
divided by the predicted gas mass as a function of stellar mass. We
find a negligible offset between the measured and predicted gas mass
measurements and a scatter of $\sigma$ = 0.361 dex, slightly higher
than found in the color-limited PGF calibration itself and slightly
higher than the probability density field model. We next show the
log(G/S) distributions predicted for RESOLVE-B with the color-limited
linear fit, where we add random scatter with $\sigma$ = 0.31 dex to
log(G/S) estimates. The second test reveals the weakness of the
color-limited approach as it cannot reproduce the distribution of
log(G/S) for red galaxies in RESOLVE-B.

%Second, below the stellar mass limit, we are
%more likely to have observed those galaxies with the highest gas
%masses, as those with lower gas masses will require the longest
%integration times.

%In panels c and d of Figure \ref{fg:testgas} we compare the predicted
%gas masses from calibrations based on two representative galaxy
%sample: the NFGS (panel c, K13) and GASS (panel d,
%\citealp{2013MNRAS.436...34C}). These two samples have fractional gas
%mass limited data like the RESOLVE survey, but their galaxy data sets
%are not statistically representative of the true galaxy population. 

In the third row, we test the calibration based on the NFGS given in
K13, \mbox{log(G/S) = $-0.98$\cujm{} + 2.70}. The NFGS has been run
through the same photometric pipeline and stellar population modeling
code as for this work, so we are able to use the same SED modeled
colors and stellar mass estimates to predict gas masses. There is a
significant offset (0.172 dex) towards underpredicting gas mass
measurements, with that offset increasing for low stellar mass
galaxies. The same issue can be seen in the second test, where the K13
calibration systematically underestimates the distribution of log(G/S)
for blue galaxies and overestimates the distribution of log(G/S) for
red galaxies in RESOLVE-B.

A similar problem is seen for the color-limited (NUV$-r$ $<$ 4.5) PGF
calibration based on GASS from \citet{2013MNRAS.436...34C},
\mbox{log(G/S) =$-0.234$log($\mu_{*}$)$-0.342$(NUV$-r$) + 2.329}. This
color-limited calibration attempts to reduce residuals for gas-rich
galaxies. We note that the GASS survey calibration does not take into
account a helium correction factor so we multiply the gas mass value
obtained from this calibration by 1.4 to compare with the gas mass
measurements in RESOLVE-B. To match their parameters we use NUV$-r$
total colors measured for the RESOLVE survey, which are similar to the
Kron NUV$-r$ colors measured for the GASS survey, although subject to
differences in the optical sky background subtraction and magnitude
extrapolation. GASS defines \mbox{$\mu_{*}$ =
  $M_{*}$/(2$\pi$$R^2_{50,z}$)}, where $R^2_{50,z}$ is the $z$-band
Petrosian half-light radius in kpc. We use the stellar mass
measurements from this work to nonetheless compute M$_{star}$ and
convert predicted G/S to predicted gas mass.\footnote{The GASS stellar
  masses are computed using the SED fitting code of
  \citet{2007ApJS..173..267S} that populates a grid of stellar
  population models with a Chabrier IMF using the
  \citet{2003MNRAS.344.1000B} stellar population synthesis
  code. Comparison between the stellar masses of
  \citet{2005ApJ...619L..39S}, \citet{2003MNRAS.341...54K},
  \citet{2009AJ....138..579K}, and K13 indicates that the stellar
  masses computed as in \citet{2007ApJS..173..267S} should be
  extremely similar to the K13 stellar masses used in this work
  (within 0.1 dex).}  Due to the constraint of requiring NUV total
magnitudes we can use only 281 galaxies for testing the GASS
calibration. The bottom row of Figure \ref{fg:testgs} shows that as
for the NFGS, there is a large offset (0.235 dex) between the
predicted and measured gas masses that increases for lower stellar
mass galaxies, in the sense that the calibration underpredicts gas
masses. Also apparent is the underestimate of the distribution of
log(G/S) for blue galaxies in RESOLVE-B.

Both the NFGS and GASS are representative galaxy samples. The NFGS is
a representative sample of galaxies designed to cover all
morphological types over a range of galaxy magnitudes in rough
proportion to the $B$-band luminosity function
\citep{2000ApJS..126..271J}. The GASS sample is designed to examine
the full range of massive galaxies with a flat distribution in stellar
mass $>$ 10$^{10}$ \msun{} \citep{2010MNRAS.403..683C}. Both samples
have complete HI data to fractional mass limits $<$ 1-5\% of the
stellar mass, making them seem like ideal calibration samples. They do
not, however, statistically represent the galaxy population of the
local universe, in particular the dominance of low-mass gas-rich
dwarfs. We note that in \citet{2012MNRAS.424.1471L}, the authors
create a calibration using GASS that includes NUV$-r$ color, stellar
mass, stellar mass surface density, and $g-i$ color gradient and more
successfully replicates gas masses for gas-rich galaxies than does the
\citet{2012A&A...544A..65C} calibration. We do not show the
\citet{2012MNRAS.424.1471L} calibration due to the large quantity of
covariant variables required, which are also difficult to calculate
from the data that we have available.

We have also tested (but do not show) the PGF calibration from
\citet{2012ApJ...756..113H} which uses the ALFALFA survey to define a
PGF calibration. This calibration relies on SDSS Petrosian $g-r$ color
and stellar mass surface density. The latter is defined in the same
way as for GASS with $z$-band Petrosian radii. We find that the
ALFALFA calibration overpredicts gas masses with an offset of $-0.19$
dex. This over prediction of gas masses is not surprising since the
ALFALFA survey is HI flux limited and weighted towards gas-rich
objects.

Lastly, we have tested (but do not show) the PGF calibrations from
\citet{2004ApJ...611L..89K} and \citet{2009MNRAS.397.1243Z}, both
based on heterogeneous samples of HI detections from the HyperLeda
database \citep{2003A&A...412...45P} crossmatched to catalogs from
SDSS and 2MASS. The \citet{2004ApJ...611L..89K} calibrations rely on
catalog \cur{} or \cuk{} color, while the \citet{2009MNRAS.397.1243Z}
calibration relies on a combination of catalog \cgr{} color and
$i$-band surface brightness. Both use stellar masses computed
following the prescription from \citet{2003ApJS..149..289B}. We mimic
these quantities, removing galaxies without quality 2MASS $K$-band
data (limiting the analysis to 182 galaxies). The PGF calibrations
from \citet{2004ApJ...611L..89K} and \citet{2009MNRAS.397.1243Z} show
very large scatter of $\sim$0.7 and $\sim$0.63 respectively, but
produce surprisingly small offsets ($-0.06$ and $-0.02$ dex
respectively), likely due to fortuitously cancelling systematics
(underrepresentation of the dwarf galaxy population but
overrepresentation of galaxies with high 21cm flux).

Based on these comparisons, we stress the importance of matching the
selection criteria and completeness of the calibration data set to the
properties of the data set for which predictions are desired. Using
PGF calibrations for a small representative galaxy sample cannot
reproduce the gas masses for a volume-limited survey, which is
dominated by low-mass, gas-rich galaxies.

The purpose of these z=0 PGF calibrations is to enable measurement of
galaxy cold gas masses for large surveys of the nearby
universe. Previous uses of PGF calibrations include identifying galaxy
mass scales connected to transitions in gas-content
(\citealp{2004ApJ...611L..89K}, K13), estimation of the HI mass
function \citep{2009MNRAS.397.1243Z}, exploration of the dependence of
the stellar mass-metallicity relationship on gas content
\citep{2009MNRAS.397.1243Z}, and galaxy clustering statistics within
galaxy populations of varying gas content \citep{2012MNRAS.424.1471L}.

These new calibrations are suited for similar studies in
volume-limited nearby galaxy surveys, for which obtaining complete HI
data is not feasible. In a companion paper (Eckert et al.\ in prep.),
we use the probability density field model to create distributions of
log(G/S) for RESOLVE and ECO galaxies, which we then combine with the
stellar mass likelihood distributions produced by the SED fitting
described in \S \ref{sec:mstar} to produce baryonic mass likelihood
distributions for each galaxy. We use Monte Carlo sampling from these
baryonic mass likelihood distributions to construct baryonic mass
functions for the RESOLVE and ECO data sets. In Florez et al. (in
prep.) the probability density field model is used to examine the gas
content of galaxies in the lowest-density environments in
ECO. Finally, we caution against the use of these calibrations at much
higher redshift as galaxy properties (including gas-content and color)
may exhibit different relationships at higher redshifts.

% The new PGF calibration may present a useful foil for
%  studying galaxy gas content at higher redshift and evaluating the
%  changing galaxy population over cosmic time.}

\section{Conclusions}
\label{sec:conclusions}

We have presented the newly reprocessed NUV$ugrizYJHK$ photometry for
the RESOLVE survey, all provided in a machine readable table (Table
\ref{tb:phottable}). We have also provided new z=0
volume-limited calibrations of the PGF technique using both linear
fits (limited to exclude the reddest colors) and, more powerfully, a
2D model of the full probability density field distribution of
log(G/S) vs.\ color (or ``modified color,'' including axial ratio). We
highlight the main conclusions of this work below:

\textbullet\ The new photometry uses improved background subtraction
and multiple flux extrapolation methods to yield brighter magnitudes,
bluer colors, and larger sizes than catalog photometry (see \S
\ref{sec:photdata} and Figures \ref{fg:photcompmagrad} and
\ref{fg:photcompcolor}). Multiple flux extrapolation techniques allow
us to measure systematic errors, key for estimating reliable stellar
masses (see \S \ref{sec:mstar}) and computing SED-modeled colors.

\textbullet\ This new photometry reveals a real increase in scatter
around the red sequence, which we attribute to the fact that our flux
extrapolation routines do not suppress color gradients, unlike the
SDSS model magnitude algorithm (see Figure \ref{fg:photcompcolor}).

\textbullet\ We provide linear PGF calibrations between log(G/S) and
color, however due to the breakdown in the relationship between
log(G/S) and color for red galaxies, these linear fits must be
color-limited and do not permit estimation of gas masses for galaxies
beyond the red color cutoff (see \S \ref{sec:simplecals} and Table
\ref{tb:simplecals}). The color-limited PGF calibrations also do not
account for the 24\% of galaxies that are confused, have weak upper
limits, or lack reliable detections, biasing the fits towards higher
gas-to-stellar mass ratios at the red end.

\textbullet\ We find that axial ratio correlates significantly with
residuals in log(G/S), independent of stellar mass, which has a
covariant correlation that is also artificially enhanced for an
approximately baryonic mass limited sample such as ours (see \S
\ref{sec:residexam} and Figures \ref{fg:gsresidall} and
\ref{fg:2dresids}). The correlation between log(G/S) residuals and
axial ratio may be at least partially related to dust and internal
extinction, which also may partially drive the PGF calibration along
with long term star formation rate.  Planar fits between color, axial
ratio, and log(G/S) provide linear combinations of color and axial
ratio (``modified color'') that yield tighter PGF calibrations than
using color alone (see \S \ref{sec:modcolpgfcal} and Table
\ref{tb:modcolorcals}).

\textbullet\ We provide a new type of PGF calibration using a 2D model
to fit the entire probability density field of log(G/S) vs.\ modified
color (see \S \ref{sec:cals} and Figures \ref{fg:goversone} and
\ref{fg:goverstwo}). Within this calibration we are able to
statistically model galaxies missing from the color-limited linear
fits.

\textbullet\ The probability density field model PGF calibration
yields log(G/S) probability distributions for individual galaxies (see
Figure \ref{fg:goversprob}). For red galaxies, for which linear fits
cannot accurately predict log(G/S), the full-probability PGF yields
two-component distributions, representing both quenched and star
forming galaxies (see Figure \ref{fg:testgs}).

\textbullet\ We test our color-limited and full-probability PGF
calibrations as well as literature calibrations using the RESOLVE-B
data set and find that our calibrations are well suited to predicting
gas masses for a volume-limited sample (see Figure \ref{fg:testgs}).
Previously published calibrations generally systematically over- or
underpredict gas masses, partly due to the different selection
criteria of the calibration samples.

We emphasize the importance of defining a calibration sample with the
same selection criteria and completeness as the sample for which
predictions are desired.  Here we have presented new PGF calibrations
using the RESOLVE survey that are ideal for application to
volume-limited data sets complete to a baryonic mass (or optical
luminosity) limit. These PGF calibrations will be used to determine
baryonic masses in a companion paper (Eckert et al.\ in prep.) that
examines the baryonic mass function as a function of environment in
RESOLVE-B and ECO .

\acknowledgements 

We would like to thank Susan Neff, Min Hubbard, Karl Forster, and the
\textit{GALEX} team for prioritizing the RESOLVE-A footprint in the
\textit{GALEX} MIS survey as well as providing us with early
\textit{GALEX} data products for this region. We would
  also like to thank our anonymous referee whose comments helped us
  refine and improve this work. We also thank Michael Blanton for
useful discussions on SDSS error statistics calculations. We thank
Jerry Sellwood, Andrew Trinker, David Hendel, Kate Storey-Fisher,
Katrina Litke, and Kirsten Hall for providing feedback to improve the
RESOLVE photometry. We are grateful to Jonathan Florez
  for testing software to compute log(G/S) distributions using the
  probability density model. We thank Chris Clemens and Andreas
Berlind for helpful comments that improved this work. K.{} Eckert,
S.{} Kannappan, D.{} Stark, M.{} Norris, E.{} Snyder, and E.{}
Hoversten were supported in this research by NSF CAREER grant
AST-0955368. K.{} Eckert and D.{} Stark acknowledge support from GAANN
Fellowships. K.{} Eckert, D.{} Stark, and A.{} Moffett acknowledge
support from North Carolina Space Grant Fellowships. D.{} Stark and
A.{} Moffett were also supported by the Royster Society Dissertation
Completion Fellowship, and A.{} Moffett was supported by the NASA
Harriet Jenkins Fellowship.

This work is based on observations from the SDSS. Funding for SDSS-III
has been provided by the Alfred P. Sloan Foundation, the Participating
Institutions, the National Science Foundation, and the U.S. Department
of Energy Office of Science. The SDSS-III web site is
http://www.sdss3.org/. SDSS-III is managed by the Astrophysical
Research Consortium for the Participating Institutions of the SDSS-III
Collaboration including the University of Arizona, the Brazilian
Participation Group, Brookhaven National Laboratory, Carnegie Mellon
University, University of Florida, the French Participation Group, the
German Participation Group, Harvard University, the Instituto de
Astrofisica de Canarias, the Michigan State/Notre Dame/JINA
Participation Group, Johns Hopkins University, Lawrence Berkeley
National Laboratory, Max Planck Institute for Astrophysics, Max Planck
Institute for Extraterrestrial Physics, New Mexico State University,
New York University, Ohio State University, Pennsylvania State
University, University of Portsmouth, Princeton University, the
Spanish Participation Group, University of Tokyo, University of Utah,
Vanderbilt University, University of Virginia, University of
Washington, and Yale University. This work is based on observations
made with the NASA Galaxy Evolution Explorer. GALEX is operated for
NASA by the California Institute of Technology under NASA contract
NAS5-98034.  This publication makes use of data products from the Two
Micron All Sky Survey, which is a joint project of the University of
Massachusetts and the Infrared Processing and Analysis
Center/California Institute of Technology, funded by the National
Aeronautics and Space Administration and the National Science
Foundation. This work is based in part on data obtained as part of the
UKIRT Infrared Deep Sky Survey.  This work uses data from the Arecibo
observatory. The Arecibo Observatory is operated by SRI International
under a cooperative agreement with the National Science Foundation
(AST-1100968), and in alliance with Ana G. Méndez-Universidad
Metropolitana, and the Universities Space Research Association. This
work is based on observations using the Green Bank Telescope. The
National Radio Astronomy Observatory is a facility of the National
Science Foundation operated under cooperative agreement by Associated
Universities, Inc. \textit{Swift} UVOT was designed and built in
collaboration between MSSL, PSU, SwRI, Swales Aerospace, and GSFC, and
was launched by NASA. We would like to thank all those involved in the
continued operation of UVOT at PSU, MSSL, and GSCF.

%\bibliographystyle{apj}
%\bibliography{/srv/one/keckert/papers/bib/paperbib}

\end{document}